\documentclass[12pt]{article}

\usepackage{pstricks}
\usepackage{epsfig}

\usepackage{amsmath}
\usepackage{amssymb}
\usepackage{amsthm}
\usepackage{array}
\usepackage{IEEEtrantools}
\usepackage{mathrsfs}
\usepackage[round, longnamesfirst]{natbib}
\usepackage{thmtools}
\usepackage{thm-restate}
\usepackage{enumitem}
\usepackage{url}
\usepackage{lipsum}
\usepackage{graphicx}
\usepackage{wrapfig}
\usepackage{subcaption}
\usepackage{mathabx}
\usepackage{comment}
\usepackage{algpseudocode}
\usepackage{bm}
\usepackage{cancel}
\usepackage{subcaption}
\usepackage[font=small]{caption}

\newcounter{theorem-counter}

\theoremstyle{definition}

\theoremstyle{plain}
\newtheorem*{theorem*}{Theorem}
\newtheorem*{corollary*}{Corollary}
\newtheorem*{lemma*}{Lemma}
\newtheorem{theorem}[theorem-counter]{Theorem}

\newtheorem{proposition}{Proposition}

\theoremstyle{definition}
\newtheorem{definition}{Definition}

\newtheorem{example}{Example}
\newtheorem*{example*}{Example}

\theoremstyle{remark}

\newtheorem{remark}{Remark}
\DeclareMathOperator{\argmax}{argmax}

\newcommand{\Tre}{\unrhd}
\newcommand{\Tr}{\rhd}

\newcommand{\ti}{\tilde}

\newcommand{\bs}[1]{\boldsymbol{#1}}
\newcommand{\tbs}[1]{\tilde{\bs{#1}}}

\newcommand{\eucp}[0]{\mathbb{R}_+^L}

\setlist{nolistsep}

\usepackage[onehalfspacing]{setspace}

\title{Revealed preference and revealed preference cycles: \\ a survey}
\date{April 2024}
\author{Pawe\l{} Dziewulski, Joshua Lanier, and John K.-H. Quah}
\begin{document}
\maketitle

\begin{abstract}
Afriat's Theorem (1967) states that a dataset can be thought of as being generated by a consumer maximizing a continuous and increasing utility function if and only if it is free of revealed preference cycles containing a strict relation.  The latter property is often known by its acronym, GARP (for {\em generalized axiom of revealed preference}).  This paper surveys extensions and applications of Afriat's seminal result.  We focus on those results where the consistency of a dataset with the maximization of a utility function satisfying some property can be characterized by a suitably modified version of GARP.
\end{abstract}



\section{Introduction}

Revealed preference analysis is built upon the observation that a utility maximizer who purchases a consumption bundle $\bs{x}$ when bundle $\bs{y}$ was cheaper is revealing that $\bs{x}$ is preferred to $\bs{y}$ (i.e.\ $\bs{x}$ yields more utility than $\bs{y}$). Thus, for instance, utility maximization prohibits a consumer from revealing that $\bs{x}$ is preferred to $\bs{y}$ (i.e.\ purchasing $\bs{x}$ when $\bs{y}$ was cheaper) and later on revealing that $\bs{y}$ is preferred to $\bs{x}$ (purchasing $\bs{y}$ when $\bs{x}$ is cheaper). This simple prohibition on cycles in the revealed preference relation is naturally extended to the \emph{generalized axiom of revealed preference} (GARP) which states that there is no sequence of bundles $\bs{x}^1, \bs{x}^2, \ldots, \bs{x}^K$ where $\bs{x}^1$ is revealed preferred to $\bs{x}^2$ which is revealed preferred to $\bs{x}^3$ and so on up to $\bs{x}^K$ and further $\bs{x}^K$ is revealed preferred to $\bs{x}^1$. 

While it is obvious that utility maximizers satisfy GARP, Afriat's Theorem (\citet{afriat67}) establishes that GARP is more than just a consequence of utility maximization, it is rather the \emph{only} consequence in the sense that whenever a consumer satisfies GARP there exists some well-behaved utility function which explains the consumer's purchases. As we shall see, GARP-style acyclicity conditions play a fundamental role in the revealed preference literature and this fundamental role is the focus of the present survey.

\paragraph{Revealed Preferences and Acyclicity.}

To elaborate further we introduce some notation. Suppose we observe a consumer purchase consumption bundle $\bs{x}^t$ (a vector with an entry reporting the quantity consumed of each good) when prices are $\bs{p}^t$ (a vector with an entry reporting the price of each good) for some $t = 1,2,\ldots, T$. We write $\bs{x}^t \ R \ \bs{x}^s$ and say that $\bs{x}^t$ is {\em revealed preferred} to $\bs{x}^s$ if the bundle $\bs{x}^s$ costs less than $\bs{x}^t$ in observation $t$. Similarly, we write $\bs{x}^t \ P \ \bs{x}^s$ and say that $\bs{x}^t$ is {\em strictly revealed preferred} to $\bs{x}^s$ if the bundle $\bs{x}^s$ costs strictly less than $\bs{x}^t$ in observation $t$. GARP then is the requirement that the revealed preference relations $R$ and $P$ do not contain cycles in the sense that there is no sequence of observations $t_1,t_2, \ldots, t_K$ so that 
\begin{equation} \label{eq:GARP-intro}
	\bs{x}^{t_1} \ R \ \bs{x}^{t_2} \ R \ \bs{x}^{t_3} \ R \ \ldots \ R \ \bs{x}^{t_K}, \text{ and } \ \bs{x}^{t_K} {P} \ \bs{x}^{t_1}.
\end{equation}
Afriat's Theorem shows that GARP characterizes well-behaved utility maximization in the sense that there exists an increasing and continuous utility function $U$ which explains the behavior of the consumer (i.e.\ in each observation $t$ the bundle chosen $\bs{x}^t$ yields more utility than any other affordable bundle) if and only if the consumer satisfies GARP. 

Many of the assumptions involved in Afriat's Theorem can be relaxed without substantially changing the result. For instance, \citet{reny15} shows that even when the dataset has infinitely many observations (whether countable or uncountable) GARP is necessary and sufficient for a dataset to be rationalized by an increasing utility function. Similarly, \citet{forges09} show that a natural extension of GARP characterizes well-behaved utility maximization when the consumer is faced with non-linear prices. \citet{polisson-quah13} show that while a utility maximizer may violate GARP when the consumption space is discrete (i.e.\ all goods can only be consumed in non-negative integer quantities) this conclusion is reversed when there is at least one (potentially unobserved) divisible good for the consumer to purchase. \citet*{nishimura-oK-quah17} show that the maximization of utility function that is increasing in a given ordering of interest (such as ordering by first order stochastic dominance over the space of simple lotteries) can be characterized by a suitable modified version of GARP.  

A related acyclicity condition known as SARP characterizes well-behaved utility maximization with single-valued demand (instead of a demand correspondence). To define SARP let us write $\bs{x}^t \ S \ \bs{x}^s$ to mean that $\bs{x}^s$ was cheaper than $\bs{x}^t$ in observation $t$ and in addition $\bs{x}^t \neq \bs{x}^s$. The consumer's behavior satisfies the strong axiom of revealed preference (SARP) if $S$ is acyclic in the sense that there is no sequence $t_1,t_2, \ldots, t_K$ so that $\bs{x}^{t_1} \ S \ \bs{x}^{t_2} \ S \ \ldots \ S \ \bs{x}^{t_K} \ S \ \bs{x}^{t_1}$. \citet{matzkin-richter91} and \citet{lee-wong05} show that SARP characterizes well-behaved utility maximization with demand functions in the sense that there exists a well-behaved utility function $U$ which generates a demand function (i.e.\ the set of consumption bundles which maximize the utility function on any given linear budget set is singleton) if and only if the consumer satisfies SARP. 

It is easy to come up with purchases which satisfy SARP and nevertheless cannot be explained as the result of a consumer maximizing a \emph{differentiable} utility function. Thus, if one wishes to explain purchasing behavior as the result of differentiable utility maximization, it is not enough that the consumer satisfies SARP.  \citet{chiappori-rochet87} and \citet{matzkin-richter91} show that if consumer satisfies SARP and in addition $\bs{p}^t \neq \bs{p}^s$ implies $\bs{x}^t \neq \bs{x}^s$ for all observations $t,s$ then there exists an increasing and infinitely differentiable utility function which generates a demand function and which explains the purchasing behavior of the consumer.

The previously mentioned results are surveyed in Sections \ref{s:garp-market} and \ref{s:garp-general}. Section \ref{s:extension} collects various results which do not fit cleanly into the previous sections. Included in this latter section is a discussion of (i) the issue of imperfect utility maximization as embodied by the cost-efficiency approach of \citet{Afriat1973}, (ii) a model of approximate utility maximization which turns out to be characterized by acyclicity of the strict revealed preference relation $P$, (iii) a model in which expenditure directly enter the utility function, and (iv) the role of acyclicity conditions in mechanism design. 


\paragraph{What we do not cover.} Our survey focuses on acyclicity conditions because these conditions play a fundamental role in the revealed preference literature and because this focus allows us to keep the paper relatively concise. The downside of this focus is that we inevitably neglect many important results in the revealed preference literature which do not involve acyclicity conditions. Here we briefly mention some of the important works which we do not cover in detail. This will at least provide the relevant references for the interested reader. We also refer the reader to the excellent book on revealed preference theory by \citet{EcheniqueChambersBook}. 

\citet{hurwicz-uzawa71} provide necessary and sufficient conditions on a demand function (as opposed to a finite set of choices) in order for this demand function to be generated by a well-behaved utility function. \citet{varian83} characterizes utility maximization where the utility function is required to be homothetic, weakly separable, or additively separable. \citet{diewert-parkan85} extend the weakly separable result in \citet{varian83}. \citet{cherchye-demuynck-derock-hjertstrand15} show how Varian's test for weak separability can be implemented as an integer programming problem.

\citet{green-srivastava86} and \citet{kubler-selden-wei14}, building on \citet{varian83}, characterize expected utility maximization with concavity and \citet{polisson-quah-renou20} characterize expected utility maximization without concavity. \citet{echeniquekota15} characterize subjective (i.e.\ the state probabilities are not given) expected utility maximization. \citet{EcheniqueImaiSaitoExperiment} characterize exponential discounted utility. For an excellent survey of the revealed preference literature for choice under risk, uncertainty, and time see \citet{echenique20}. 

\citet{browning89} characterizes additive across-period utility maximization (i.e.\ the utility function takes the form $\sum_{t=1}^T U( \tbs{x}^t )$ where $\tbs{x}^t$ is consumption in period $t$). Interestingly, it is shown in \citet{brown-calsamiglia07} that the same condition found in \citet{browning89} also characterizes quasilinear utility maximization. Quasilinear utility maximization is further investigated in \citet{castillo-freer23} and \citet{allen-rehbeck21}.  

There is also a related literature covering testing in economic equilibrium models which is surveyed in \citet{carvajal24}. \citet{BrownMatzkin1996} study testable restrictions in an exchange economy. This analysis is extended in \cite{BrownShanon2000} who study the testable restriction of competitive equilibria, and \cite{Carvajal2010} who extends the analysis to competitive economies with externalities. \cite{Belgians2007} and \cite{Belgians2010} characterize the behavioral consequences of the collective household model presented in \citet{BrowningChiappori1998}. \citet{carvajal-deb-fenske-quah13} characterize the testable implications of Cournot competition. 

Finally, there is a literature on testing random utility models in budgetary environments. See \citet{mcfadden-richter91}, \citet{mcfadden05}, \citet{hoderlein-stoye14}, \citet{hoderlein15}, \citet{kawaguchi17}, and \citet{kitamura-stoye18}.

\section{Rationalizable behavior in a market setting}\label{s:garp-market}

\subsection{Afriat's Theorem} \label{ss:afriat}

We begin our discussion with Afriat's Theorem. To set the stage, we assume that there are $L$ goods that can be purchased in any non-negative quantity and a consumer who purchases a bundle of the $L$ goods $\bs{x} = (x_1,x_2, \ldots, x_L) \in \mathbb{R}_+^L$, where $x_{\ell}$ is the quantity of good $\ell$ consumed. A \emph{preference relation} $\succeq$ is a complete and transitive binary relation.\footnote{The binary relation $\succeq$ is complete if $\bs{x} \succeq \bs{y}$ or $\bs{y} \succeq \bs{x}$ for all $\bs{x},\bs{y}$ and $\succeq$ is transitive if $\bs{x} \succeq \bs{y}$ and $\bs{y} \succeq \bs{z}$ implies $\bs{x} \succeq \bs{z}$ for all $\bs{x}, \bs{y}, \bs{z}$.} A \emph{utility function} $U$ is any function that maps the consumption bundles $\mathbb{R}_{+}^L$ into the real line $\mathbb{R}$. Each utility function $U$ \emph{generates} a preference relation $\succeq$ by defining $\succeq$ so that $U(\bs{x}) \geq U(\bs{y})$ if and only if $\bs{x} \succeq \bs{y}$ for all $\bs{x}, \bs{y}$. We next discuss the observables.

We observe a \emph{purchase dataset} of the form $D = \big\{( \bs{x}^t, \bs{p}^t )\big\}_{t \leq T}$ where $\bs{x}^t \in \mathbb{R}_+^L$ denotes the consumption bundle selected by the consumer in period $t$ under the prevailing prices $\bs{p}^t = (p_1^t, p_2^t,\ldots, p_L^T) \in \mathbb{R}_{++}^L$ where $p_{\ell}^t$ is the price of good $\ell$ in period $t$. We wish to know if the consumer selected the consumption bundles in $D$ to maximize utility. More formally, we wish to know if $D$ can be \emph{rationalized} by a utility function $U$ in the sense that, for all $t$, the bundle chosen $\bs{x}^t$ yields more utility than any other affordable bundle, i.e.\
\begin{equation}\label{eq:ratio}
	\bs{p}^t \cdot \bs{x}^t \geq \bs{p}^t \cdot \bs{y} \ \text{ implies } \ U(\bs{x}^t)  \geq  U(\bs{y}).
\end{equation}
A more modest objective is to determine if there exists a preference relation $\succeq$ which \emph{rationalizes} the data $D$ in the sense that, for all $t$, the bundle chosen $\bs{x}^t$ is preferred to any other affordable bundle, i.e.\
\begin{equation*}\label{eq:ratio2}
	\bs{p}^t \cdot \bs{x}^t \geq \bs{p}^t \cdot \bs{y} \ \text{ implies } \ \bs{x}^t  \succeq  \bs{y}.
\end{equation*}
If $U$ rationalizes $D$ then the preference relation $\succeq$ generated by $U$ also rationalizes $D$ and thus the concept of a utility rationalization is more demanding than that of a preference rationalization.

Without additional restrictions there are no testable implications of utility or preference maximization. In particular, every dataset $D$ can be rationalized by (i) the preference relation in which the consumer is indifferent between all consumption bundles and (ii) a constant utility function. Thus, we ought to entertain some additional properties of preference relations and utility functions.

Here are some properties we might like to impose on our rationalizing preference relations and utility functions. A preference relation $\succeq$ is \emph{locally non-satiated} if, for every bundle $\bs{x} \in \mathbb{R}_+^L$ and every neighborhood $N$ of $\bs{x}$ there exists some $\bs{y} \in N$ so that $\bs{y} \succ \bs{x}$ (where $\bs{y} \succ \bs{x}$ means $\bs{y} \succeq \bs{x}$ and $\bs{x} \not\succeq \bs{y}$). The function $U: \mathbb{R}_+^L \rightarrow \mathbb{R}$ is \emph{increasing} if $\bs{x} \geq \bs{y}$ implies $U(\bs{x}) \geq U(\bs{y})$ and $\bs{x} \gg \bs{y}$ implies $U(\bs{x}) > U(\bs{y})$. The function $U: \mathbb{R}_+^L \rightarrow \mathbb{R}$ is \emph{strictly increasing} if $\bs{x} > \bs{y}$ implies $U(\bs{x}) > U(\bs{y})$.\footnote{As usual, for $\bs{x} = (x_1,x_2, \ldots, x_L)$ and $\bs{y} = (y_1, y_2, \ldots, y_L)$ we write $\bs{x} \geq \bs{y}$ to mean $x_{\ell} \geq y_{\ell}$ for all $\ell$, we write $\bs{x} > \bs{y}$ to mean $\bs{x} \geq \bs{y}$ and $\bs{y} \not\geq \bs{x}$, and we write $\bs{x} \gg \bs{y}$ to mean $x_{\ell} > y_{\ell}$ for all $\ell$.} A strictly increasing utility function obviously generates a locally non-satiated preference relation and thus a dataset which can be rationalized by a strictly increasing utility function can also be rationalized by a locally non-satiated preference relation. 

\begin{figure}[t]
	\centering
	\begin{center}
		\includegraphics[scale = 0.4]{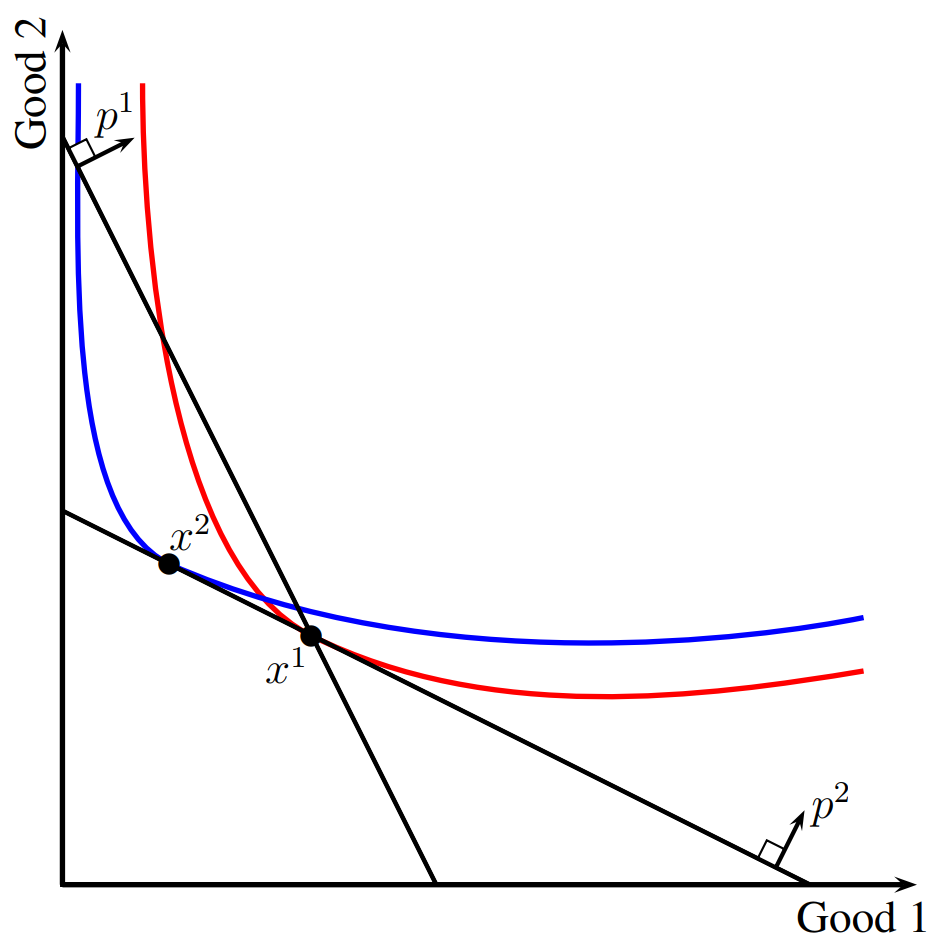}
	\end{center}
	\vspace{-3pt}

	\caption{Violation of GARP.}
	\label{fig:01}
\end{figure}

While every purchase dataset can be rationalized by a constant utility function it is easy to construct datasets which cannot be rationalized by strictly increasing utility functions. For example, let $D = \{ (\bs{x}^t, \bs{p}^t) \}_{t \leq 2}$ be the two observation dataset depicted in Figure \ref{fig:01}. We see that $\bs{x}^1$ is weakly cheaper than $\bs{x}^2$ in period 2 (i.e.\ $\bs{p}^2 \cdot \bs{x}^2 \geq \bs{p}^2 \cdot \bs{x}^1$) and $\bs{x}^2$ is strictly cheaper than $\bs{x}^1$ in period 2 (i.e.\ $\bs{p}^1 \cdot \bs{x}^1 > \bs{p}^1 \cdot \bs{x}^2)$. Any strictly increasing utility function $U$ which rationalizes $D$ must, because of the second observation, satisfy $U(\bs{x}^2) \geq U(\bs{x}^1)$ and must, because of the first observation, satisfy $U(\bs{x}^1) > U(\bs{x}^2)$. These two requirements are clearly incompatible. Another way of showing that this dataset cannot be rationalized by a strictly increasing utility function is to note that any strictly increasing utility function which rationalizes this dataset would have indifference curves tangent to the budget lines at the selected bundles. However, as shown in the figure, any such indifference curves would have to cross, which is not allowed by any increasing utility function.

From the previous discussion the following facts are immediate; (i) if a consumer buys bundle $\bs{x}^t$ when $\bs{y}$ is weakly cheaper then it must be that $U(\bs{x}^t) \geq U(\bs{y})$ and (ii) if a consumer buys bundle $\bs{x}^t$ when $\bs{y}$ is strictly cheaper then it must be that $U(\bs{x}^t) > U(\bs{y})$. This serves as the motivation for the following revealed preference relations. We write $\bs{x}^t \ R \ \bs{y}$ and say that $\bs{x}^t$ is \emph{revealed preferred} to $\bs{y}$ if $\bs{y}$ is weakly cheaper than $\bs{x}^t$ in period $t$, i.e.\ $\bs{p}^t \cdot \bs{x}^t \geq \bs{p}^t \cdot \bs{y}$ and further we write $\bs{x}^t \ P \ \bs{y}$ and say that $\bs{x}^t$ is \emph{strictly revealed preferred} to $\bs{y}$ if $\bs{y}$ is strictly cheaper than $\bs{x}^t$ in period $t$, i.e.\ $\bs{p}^t \cdot \bs{x}^t > \bs{p}^t \cdot \bs{y}$. From (i) and (ii) above we see that, every strictly increasing utility function $U$ which rationalizes $D$ must satisfy
\begin{equation} \label{eq:rp}
	\bs{x}^t \ R \ \bs{y} \ \implies \ U(\bs{x}^t) \geq U(\bs{y}) \qquad \text{ and } \qquad \bs{x}^t \ P \ \bs{y} \ \implies \ U(\bs{x}^t) > U(\bs{y})
\end{equation}
From \eqref{eq:rp} it is clear that if $D$ can be rationalized by a strictly increasing utility function then the revealed preference relations $(R,P)$ must be acylic in the following sense. 
\begin{definition}[GARP] \label{def:GARP1}
	The purchase dataset $D = \big\{( \bs{x}^t, \bs{p}^t )\big\}_{t \leq T}$ satisfies \emph{the generalized axiom of revealed preference} (GARP) if there are no sequences $t_1,t_2,\ldots, t_K$ so that
	\begin{equation} \label{eq:GARP}
		\bs{x}^{t_1} \ R \ \bs{x}^{t_2} \ R \ \bs{x}^{t_3} \ R \ \ldots \ R \ \bs{x}^{t_K}, \text{ and } \ \bs{x}^{t_K} {P} \ \bs{x}^{t_1}.
	\end{equation}
\end{definition}
To see that utility maximizers with strictly increasing utility functions always satisfy GARP let us suppose that $D$ is rationalized by a strictly increasing utility function $U$ and suppose there is some sequence $t_1, t_2,\ldots, t_K$ so that \eqref{eq:GARP} holds (i.e.\ GARP is violated). From \eqref{eq:rp} we have $U(\bs{x}^{t_1}) \geq U(\bs{x}^{t_2}) \geq \ldots \geq U(\bs{x}^{t_K}) > U(\bs{x}^{t_1})$ and so we obtain the absurdity $U(\bs{x}^{t_1}) > U(\bs{x}^{t_1})$ and the claim is proved. Surprisingly, GARP is also sufficient. The following is Afriat's Theorem from \citet{afriat67} (see also \citet{diewert73} and \cite{varian82}).\footnote{Recall that a function $U: \mathbb{R}_{+}^L \rightarrow \mathbb{R}$ is concave if $U(\alpha \bs{x} + (1-\alpha) \bs{y}) \geq \alpha U(\bs{x}) + (1-\alpha) U(\bs{y})$ for all $\bs{x},\bs{y}$ and $\alpha \in [0,1]$.}

\begin{theorem}[\citealp{afriat67}] \label{theorem:afriat}
	For any purchase dataset $D = \big\{(\bs{x}^t, \bs{p}^t)\big\}_{t \leq T}$, the following statements are equivalent.
	\begin{enumerate}
		\item The dataset $D$ can be rationalized by a locally non-satiated preference relation.
		\item The dataset $D$ satisfies GARP.
		\item There are numbers $\{u^t\}_{t \leq T}$ and strictly positive numbers $\{\lambda^t\}_{t \leq T}$ such that
		\begin{equation} \label{eq:afriat-ineqs}
			u^s \ \leq \ u^t + \lambda^t \bs{p}^t \cdot ( \bs{x}^s - \bs{x}^t ), \qquad \text{ for all } s,t \in \{1,2,\ldots, T\}.
		\end{equation}
		\item The data $D$ can be rationalized by a continuous, concave, and strictly increasing utility function.
	\end{enumerate}
\end{theorem}
The implication 1 $\Rightarrow$ 4 shows that any dataset which can be rationalized by a locally non-satiated preference relation can also be rationalized by a continuous, concave, and strictly increasing utility function. Thus, there are no testable implications of utility maximization by a continuous, concave, and strictly increasing utility function beyond what is already implied by the maximization of a locally non-satiated preference relation. Statement 3 establishes that GARP can be verified by solving a simple linear programming problem and thus determining if a given dataset satisfies GARP is computationally straightforward.

Proofs of Afriat's Theorem may be found in \cite{afriat67}, \cite{varian82}, \cite{FostelScarfTodd2004}, and \cite{PolissonRenou2016}. Below we provide our own self-contained proof which is inspired by the proof in \cite{QUAH2014}. Showing 4 $\Rightarrow$ 1 and 1 $\Rightarrow$ 2 in Afriat's Theorem is trivial. To show that 3 $\Rightarrow$ 4 we assume that we have found some numbers $\{u^t\}_{t \leq T}$ and strictly positive numbers $\{\lambda^t\}_{t \leq T}$ which satisfy \eqref{eq:afriat-ineqs} and then define $U$ by $U(\bs{x}) = \min_t u^t + \lambda^t \bs{p}^t \cdot (\bs{x} - \bs{x}^t)$. It happens that this $U$ satisfies the properties required in Statement 4. Thus, the main difficulty in proving Afriat's Theorem lies in showing 2 $\Rightarrow$ 3. The approach we take is to define the numbers $\{u^t\}_{t \leq T}$ and strictly positive numbers $\{\lambda^t\}_{t \leq T}$ recursively (in other words we define $u^1$ then we define $\lambda^1$ as a function of $u^1$ and then we define $u^2$ as a function of $u^1$ and $\lambda^1$ and then we define $\lambda^2$ as a function of all the previously defined numbers, and so forth). As we shall see, an appropriately specified recursive formula is enough to produce the desired numbers.

\begin{proof}[Proof of Afriat's Theorem.]
	As every strictly increasing utility function generates a locally non-satiated preference relation it is clear that 4 $\Rightarrow$ 1. We next show 1 $\Rightarrow$ 2. Suppose $D$ is rationalized by locally non-satiated preference relation $\succeq$. As $\succeq$ rationalizes $D$ we see that $\bs{p}^t \cdot \bs{x}^t \geq \bs{p}^t \cdot \bs{y}$ implies $\bs{x}^t \succeq \bs{y}$. Further, if $\bs{p}^t \cdot \bs{x}^t > \bs{p}^t \cdot \bs{y}$ then there exists some neighborhood $N$ of $\bs{y}$ so that $\bs{p}^t \cdot \bs{x}^t > \bs{p}^t \cdot \bs{z}$ for all $\bs{z} \in N$. As $\succeq$ is locally non-satiated there exists $\bs{z} \in N$ so that $\bs{z} \succ \bs{y}$ and thus, as $\succeq$ rationalizes $D$, we see that $\bs{x}^t \succeq \bs{z} \succ \bs{y}$ and so we have seen that $\bs{x}^t \ R \ \bs{y}$ implies $\bs{x}^t \succeq \bs{y}$ and $\bs{x}^t \ P \ \bs{y}$ implies $\bs{x}^t \succ \bs{y}$ for all $t$ and $\bs{y}$. As $\succeq$ is transitive it is now obvious that GARP holds and so we have shown that 1 $\Rightarrow$ 2. 
	
	We next show 2 $\Rightarrow$ 3. We write $\bs{x}^t \ R^* \ \bs{x}^s$ if there exists $t_1, t_2, \ldots,t_K$ so that $\bs{x}^{t_1} \ R \ \bs{x}^{t_2} \ R \ \ldots \ R \ \bs{x}^{t_K} \ R\ \bs{x}^s$ (so $R^*$ is the transitive closure of $R$). Let $R_S^*$ denote the strict part of $R^*$ (that is, $\bs{x}^t \ R_S^* \ \bs{x}^s$ means $\bs{x}^t \ R^* \ \bs{x}^s$ and $\bs{x}^s \ \cancel{R^*} \ \bs{x}^t$). Let $|t|$ be the number of elements in the longest sequence $t_1, t_2, \ldots, t_K$ satisfying $\bs{x}^{t_1} \ R_S^* \ \bs{x}^{t_2} \ R_S^* \ \ldots \ R_S^* \ \bs{x}^{t_K} \ R_S^* \ \bs{x}^t$ (so, $|t|$ is the number of elements in the longest path beginning at $\bs{x}^t$ in the $R_S^*$ ordering). Reorder the observations if needed so that $|1| \leq |2| \leq |3| \leq \ldots \leq |T|$. We define $(u^t,\lambda^t)_{t \leq T}$ recursively via
	\begin{IEEEeqnarray}{rCl}
		u^i & = & \min \left( \Big\{ u^j + \lambda^j \bs{p}^j \cdot (\bs{x}^k - \bs{x}^j): \forall j,k \text{ s.t. } |j| < |i| \text{ and } |k| = |i| \Big\} \cup \{0\} \right) \label{eq:u-def2} \\
		\lambda^i & = & \max \left( \Big\{ \dfrac{ u^j - u^i }{ \bs{p}^i \cdot (\bs{x}^j - \bs{x}^i) }: \forall j \text{ s.t. } |j| < |i| \Big\} \cup \{ 1 \} \right) \label{eq:lambda-def2}
	\end{IEEEeqnarray}
	Note that $u^i$ in \eqref{eq:u-def2} only depends on $u^j$ and $\lambda^j$ for $|j| < |i|$ and thus $u^i$ only depends on $j < i$. Similarly, $\lambda^i$ in \eqref{eq:lambda-def2} only depends on $u^i$ and $u^j$ where $|j| < |i|$ and thus the numbers $(u^t,\lambda^t)_{t \leq T}$ are indeed define recursively. Note that $|j| < |i|$ implies that $\bs{x}^i \ \cancel{R} \ \bs{x}^j$ and thus $\bs{p}^i \cdot (\bs{x}^j-\bs{x}^i) > 0$ and so \eqref{eq:lambda-def2} never involves division by $0$. From these remarks it is clear that the numbers $(u^t,\lambda^t)_{t \leq T}$ are well-defined. Next, note that $\lambda^i \geq 1$ and thus $\lambda^i$ satisfies the requirement of being strictly positive. We next confirm that the numbers $(u^t,\lambda^t)_{t \leq T}$ satisfy \eqref{eq:afriat-ineqs}. 
	
	Let $s,t \in \{1,2,\ldots, T\}$. If $|t| < |s|$ then \eqref{eq:afriat-ineqs} follows immediately from \eqref{eq:u-def2} (take $i = s$). If $|t| = |s|$ then it is obvious that $u^s = u^t$ and because $D$ satisfies GARP we have $\bs{p}^t \cdot (\bs{x}^s - \bs{x}^t) \geq 0$ (this is in fact the only time GARP is invoked) and so \eqref{eq:afriat-ineqs} holds. Finally, if $|s| < |t|$ then (using \eqref{eq:lambda-def2} with $i = t$)
	\begin{equation*}
		u^t + \lambda^t \bs{p}^t \cdot ( \bs{x}^s - \bs{x}^t ) \geq u^t + \dfrac{ u^s - u^t }{ \bs{p}^t \cdot ( \bs{x}^s - \bs{x}^t ) } ( \bs{p}^t \cdot ( \bs{x}^s - \bs{x}^t )) = u^s
	\end{equation*}
	and so the numbers $(u^t,\lambda^t)_{t \leq T}$ satisfy \eqref{eq:afriat-ineqs} and thus we have proved 2 $\Rightarrow$ 3.
	
	To complete the proof we need to show 3 $\Rightarrow$ 4. Take any numbers $u^t, \lambda^t$, for $t \leq T$, which satisfy \eqref{eq:afriat-ineqs} and define $U:\mathbb{R}_+^L \to \mathbb{R}$ by
	\begin{equation}\label{eq:utility-afriat}
		U(\bs{x}) \ := \ \min_{t} \Big\{u^t + \lambda^t p^t\cdot (\bs{x} - \bs{x}^t) \Big\},
	\end{equation}
	which is continuous, concave, and strictly increasing as it is the pointwise minimum of a finite number of continuous, concave, and strictly increasing functions. It remains to show that $U$ rationalizes $D$. Take any $t \leq T$ and $\bs{y}$ such that $\bs{p}^t \cdot \bs{y} \leq \bs{p}^t \cdot \bs{x}^t$ or, equivalently,  $\bs{p}^t \cdot (\bs{y} - \bs{x}^t) \leq 0$. Then,
	\[
	U(\bs{y}) \ \leq \ u^t + \lambda^t p^t\cdot (\bs{y} - \bs{x}^t) \ \leq \ u^t \ \leq \ \min_{s \in T} \Big\{u_s + \lambda_s p^s\cdot (\bs{x}^t - \bs{x}^s) \Big\} \ = \ U(\bs{x}^t).
	\]
	This concludes our proof of Theorem \ref{theorem:afriat}.
\end{proof}

We conclude this subsection with some remarks.
\begin{remark}
	Let us say that the dataset $D = \{ (\bs{x}^t, \bs{p}^t) \}_{t \leq T}$ satisfies the \emph{weak axiom of revealed preferences} (WARP) if there are no observations $s$ and $t$ so that $\bs{x}^t \ R \ \bs{x}^s$ and $\bs{x}^s \ P \ \bs{x}^t$. While it is clear that a dataset which satisfies GARP also satisfies WARP surprisingly, as shown by \citet{Rose1958} and \citet{banerjee-murphy06}, when there are only two goods (so $L=2$) the reverse implication also holds. That is, when $L=2$ GARP and WARP are equivalent and so, under these conditions, we can confirm that the dataset $D$ can be rationalized by a strictly increasing, continuous, and concave utility function by checking that $D$ satisfies WARP. 
\end{remark}

\begin{remark}
	\citet{kitamura-stoye18}, building on \citet{mcfadden-richter91} (see also \citet{mcfadden05}), characterize random utility maximization when we observe the distribution of demand over a finite number of linear budget sets. That is, suppose there are $T$ sets $B^1, B^2, \ldots, B^T \subseteq \mathbb{R}_{+}^L$ where, for each $t$, we have $B^t = \{ \bs{x} \in \mathbb{R}_+^L: \bs{p}^t \cdot \bs{x} \leq m^t \}$ for some price vector $\bs{p}^t \in \mathbb{R}_{++}^L$ and some expenditure level $m^t > 0$. For each $t$, let $F^t$ denote a probability measure concentrated on $B^t$. A random utility model (RUM) is a distribution $F$ over strictly increasing utility functions. The RUM $F$ \emph{rationalizes} $D = \{ (F^t, B^t) \}_{t \leq T}$ if, for each $t$ and for all Borel sets $A \subseteq \mathbb{R}_{+}^L$ the probability $F^t(A)$ is equal to the probability with which a utility function $U$ drawn from $F$ has an optimum over $B^t$ which is inside $A$. \citet{kitamura-stoye18} provide necessary and sufficient conditions for $D$ to be rationalized by a RUM. 
\end{remark}

\begin{remark}
	\citet{Richter1966} characterizes preference maximization using a stronger notion of what it means for a preference to rationalize behavior. Let $\mathcal{B}$ be some collection of non-empty subsets of $\mathbb{R}_{+}^L$ and let $c$ be a correspondence with domain $\mathcal{B}$ which satisfies $c(B) \subseteq B$ for each $B \in \mathcal{B}$. Intuitively, $c(B)$ represents the choices which the consumer finds acceptable when confronted with constraint set $B$. Let us say that the preference relation $\succeq$ \emph{strictly rationalizes} $c$ if, for all $B \in \mathcal{B}$, we have $c(B) = \{ \bs{x} \in B: \bs{x} \succeq \bs{y} \text{ for all } \bs{y} \in B \}$. In other words, in order for $\succeq$ to strictly rationalize $c$ we require that the set of bundles chosen $c(B)$ are exactly the best bundles in $B$ according $\succeq$. 
	
	Let us write $\bs{x} \ R \ \bs{y}$ if there exists $B \in \mathcal{B}$ such that $\bs{y} \in B$ and $\bs{x} \in c(B)$ (in other words, $\bs{x}$ was chosen when $\bs{y}$ was available). Also, write $\bs{x} \ P \ \bs{y}$ if there exists $B \in \mathcal{B}$ such that $\bs{y} \in B$, $\bs{x} \in c(B)$, and $\bs{y} \notin c(B)$ (in other words, $\bs{x}$ was chosen and $\bs{y}$ was not chosen). The choice correspondence $c$ satisfies the \emph{congruence axiom} if, there is no sequence $\bs{x}^1, \bs{x}^2, \ldots, \bs{x}^K$ such that
	\begin{equation*}
		\bs{x}^1 \ R \ \bs{x}^2 \ R \ \bs{x}^3 \ R \ \ldots \ R \ \bs{x}^K, \text{ and } \bs{x}^K \ P \ \bs{x}^1
	\end{equation*}
	It is easy to see that the congruence axiom is necessary for $c$ to be strictly rationalized by a preference relation. \citet{Richter1966} shows that the congruence axiom is also sufficient. In fact, Richter's result holds for any consumption space (not just the space of consumption bundles $\mathbb{R}_+^L$) and, in this respect, the result is similar to the result of \citet{nishimura-oK-quah17} which we present below as Theorem \ref{thm:NOQ1}.
\end{remark}

\subsection{SARP and demand functions} \label{ss:sarp}

The utility function supplied by Afriat's Theorem does not in general generate a single-valued demand function. To be specific, let $h(\bs{p}, m)$ be the demand generated by utility function $U$ with the constraint of spending no more than $m$ when prices are $\bs{p}$, i.e.\
\begin{equation} \label{eq:demand}
	h(\bs{p}, m) = \underset{\bs{x} \in B(\bs{p}, m) }{\argmax} \ U(\bs{x})
\end{equation}
where $B(\bs{p}, m)$ is the budget set $\{ \bs{x} \in \mathbb{R}_{+}^L: m \geq \bs{p} \cdot \bs{x} \}$. Even when $U$ is a continuous, concave, and strictly increasing utility function (these being the properties possessed by the utility function supplied by Statement 4 in Afriat's Theorem), the demand  $h(\bs{p},m)$ may be non-singleton for some values of $(\bs{p},m)$. Here we investigate how to test for utility maximization which generates demand functions (in other words, $h(\bs{p},m)$ is a singleton for all $(\bs{p},m) \in \mathbb{R}_{++}^L \times \mathbb{R}_{++}$). 

We can also define the demand $h$ generated by a preference relation. Specifically, the preference relation $\succeq$ generates the demand $h$ defined by
\begin{equation}
	h(\bs{p}, m) = \Big\{ \bs{x} \in B(\bs{p}, m): \bs{x} \succeq \bs{y} \text{ for all } \bs{y} \in B(\bs{p},m) \Big\}
\end{equation}
We shall also be interested in conditions under which the data can be rationalized by a preference relation which generates a demand function. 

If the purchase data $D= \{(\bs{x}^t, \bs{p}^t)\}_{t \leq T}$ is rationalized by a utility function $U$ which generates a demand function then clearly, for each $t$, the utility derived from $\bs{x}^t$ strictly exceeds the utility derived from any other affordable bundle. This suggests introducing the revealed preference relation $S$ where $\bs{x}^t \ S \ \bs{y}$ means that $\bs{y}$ is some affordable bundle distinct from $\bs{x}^t$ (i.e.\ $\bs{p}^t \cdot \bs{x}^t \geq \bs{p}^t \cdot \bs{y}$ and $\bs{x}^t \neq \bs{y}$). Clearly,
\begin{equation} \label{eq:rp2}
	\bs{x}^t \ S \ \bs{y} \qquad \implies \qquad U(\bs{x}^t) > U(\bs{y})
\end{equation}
where $U$ is any utility function which rationalizes the data and which generates a demand function. Notice that the relation $S$ is almost equivalent to the $R$ defined in the previous subsection, but applies only to bundles different from $\bs{x}^t$, i.e., we have $\bs{x}^t \ R \ \bs{x}^t$, but \emph{not} $\bs{x}^t \ S \ \bs{x}^t$. From \eqref{eq:rp2} it is clear that if $D$ can be rationalized by a utility function which generates a demand function then the relation $S$ must be acyclic in the following sense. 
\begin{definition}[SARP]\label{def:SARP1}
	The purchase dataset $D = \big\{(\bs{x}^t, \bs{p}^t)\big\}_{t \leq T}$ satisfies the \emph{strong axiom of revealed preferences} (SARP) if, there is no sequence $t_1, t_2, \ldots, t_K$ so that
	\begin{equation} \label{eq:SARP}
		\bs{x}^{t_1} \ S \ \bs{x}^{t_2} \ S \ \bs{x}^{t_3} \ S \ \ldots \ S \ \bs{x}^{t_K}, \text{ and } \  \bs{x}^{t_K} \ S \ \bs{x}^{t_1}
	\end{equation}
\end{definition}
While SARP is obviously a necessary condition in order for the data $D$ to be rationalized by a utility function which generates a demand function it happens that SARP is also sufficient. A version of the following theorem was originally stated in \citet{matzkin-richter91} and later improved upon in \citet{lee-wong05}.\footnote{Recall that a function $U: \mathbb{R}_{+}^L \rightarrow \mathbb{R}$ is strictly concave if $U( \alpha \bs{x} + (1-\alpha) \bs{y} ) > \alpha U(\bs{x}) + (1-\alpha) U(\bs{y})$ for all $\bs{x},\bs{y}$ and $\alpha \in (0,1)$.}
\begin{theorem}[\citealp{matzkin-richter91} and \citealp{lee-wong05}] \label{theorem:sarp}
	For any purchase dataset $D = (\bs{x}^t, \bs{p}^t)_{t \leq T}$, the following statements are equivalent.
	\begin{enumerate}
		\item $D$ can be rationalized by a preference relation which generates a demand function.
		\item $D$ satisfies SARP.
		\item There are numbers $\{u^t\}_{t \leq T}$ satisfying $\bs{x}^t = \bs{x}^s$  implies $u^t = u^s$ for all $s,t$ and strictly positive numbers $\{\lambda^t\}_{t \leq T}$ such that
		\begin{equation} \label{eq:afriat-ineqs2}
			u^s \ < \ u^t + \lambda^t \bs{p}^t \cdot ( \bs{x}^s - \bs{x}^t ),  \text{ for all } s,t \text{ such that } \bs{x}^t \neq \bs{x}^s.
		\end{equation}
		\item $D$ can be rationalized by a continuous, strictly concave, and strictly increasing utility function $U$ which generates an infinitely differentiable demand function.
	\end{enumerate}
\end{theorem}




Here we provide a sketch of the proof of Theorem \ref{theorem:sarp}.   Implications 4 $\Rightarrow$ 1 and 1 $\Rightarrow$ 2 are obvious. To show 2 $\Rightarrow$ 3 we re-purpose the proof from Section \ref{ss:afriat}. Indeed, let $|t|$ be defined as in this proof and define numbers $\{u^t\}_{t \leq T}$ and $\{\lambda^t\}_{t \leq T}$ by
\begin{IEEEeqnarray}{rCl}
	u^i & = & \min \left( \Big\{ u^j + \lambda^j \bs{p}^j \cdot (\bs{x}^k - \bs{x}^j) - 1: \forall j,k \text{ s.t. } |j| < |i| \text{ and } |k| = |i| \Big\} \cup \{0\} \right) \label{eq:u-def3} \\
	\lambda^i & = & \max \left( \Big\{ \dfrac{ u^j - u^i + 1 }{ \bs{p}^i \cdot (\bs{x}^j - \bs{x}^i) }: \forall j \text{ s.t. } |j| < |i| \Big\} \cup \{ 1 \} \right) \label{eq:lambda-def3}
\end{IEEEeqnarray}
The numbers $u^i$ and $\lambda^i$ can be shown to be well-defined using the arguments we used to show that the numbers defined in \eqref{eq:u-def2} and \eqref{eq:lambda-def2} are well-defined. Also, note that $\lambda^i \geq 1$ for all $i$ and so to finish our proof of 2 $\Rightarrow$ 3 we must verify that $\{u^t\}_{t \leq T}$ and $\{\lambda^t\}_{t \leq T}$ satisfy \eqref{eq:afriat-ineqs2}. 

Let $s,t \in \{1,2,\ldots, T\}$. If $|t| < |s|$ then \eqref{eq:afriat-ineqs2} follows immediately from \eqref{eq:u-def3} (take $i = s$). If $|t| = |s|$ then it is obvious that $u^s = u^t$ and, if further $\bs{x}^t \neq \bs{x}^s$ then, because $D$ satisfies SARP, it must be that $\bs{x}^t \ \cancel{R} \ \bs{x}^s$ and therefore $\bs{p}^t \cdot (\bs{x}^s - \bs{x}^t) > 0$ and so \eqref{eq:afriat-ineqs2} holds. Finally, if $|s| < |t|$ then (using \eqref{eq:lambda-def3} with $i = t$)
\begin{equation*}
	u^t + \lambda^t \bs{p}^t \cdot ( \bs{x}^s - \bs{x}^t ) \geq u^t + \dfrac{ u^s - u^t + 1 }{ \bs{p}^t \cdot ( \bs{x}^s - \bs{x}^t ) } ( \bs{p}^t \cdot ( \bs{x}^s - \bs{x}^t )) = u^s + 1 > u^s
\end{equation*}
and so the numbers $(u^i,\lambda^i)$ satisfy \eqref{eq:afriat-ineqs2} and thus we have proved 2 $\Rightarrow$ 3.

To show 3 $\Rightarrow$ 4 we follow the arguments in \citet{matzkin-richter91}. The utility $U$ can be constructed as follows. First, take any function $g:\mathbb{R}_+^L \to \mathbb{R}_+$ that is strictly convex and differentiable with $\partial g/ \partial x_i < 1$, for all $i = 1, \ldots, L$, and satisfies $g(\bs{x}) = 0$ if, and only if, $\bs{x} = 0$.\footnote{ %
	For, example $g(\bs{x}) := \sqrt{x_1^2 + x_2^2 + \ldots + x_L^2 + T} - \sqrt{T}$, for some $T > 0$.
}
There is a sufficiently small number $\varepsilon >0$ such that the function
\[
U(\bs{x}) \ := \ \min_{t \in T} \Big\{u^t + \lambda^t p^t\cdot (\bs{x} - \bs{x}^t) - \epsilon g(\bs{x} - \bs{x}^t) \Big\},
\]
is continuous, strictly concave, strictly increasing, and rationalizes the data. Since $U$ is continuous and strictly concave this utility function generates a demand function. This demand function will not in general be infinitely differentiable and so we fall short of delivering the utility function promised in statement 4. For the complete proof of 3 $\Rightarrow$ 4 we refer the reader to \citet{lee-wong05} where a more complex construction is used to produce the desired utility function.

\subsection{Rationalizability with a differentiable utility} \label{ss:diff}

\begin{figure}[t]
	\centering
	\begin{center}
	 	\includegraphics[scale = 0.4]{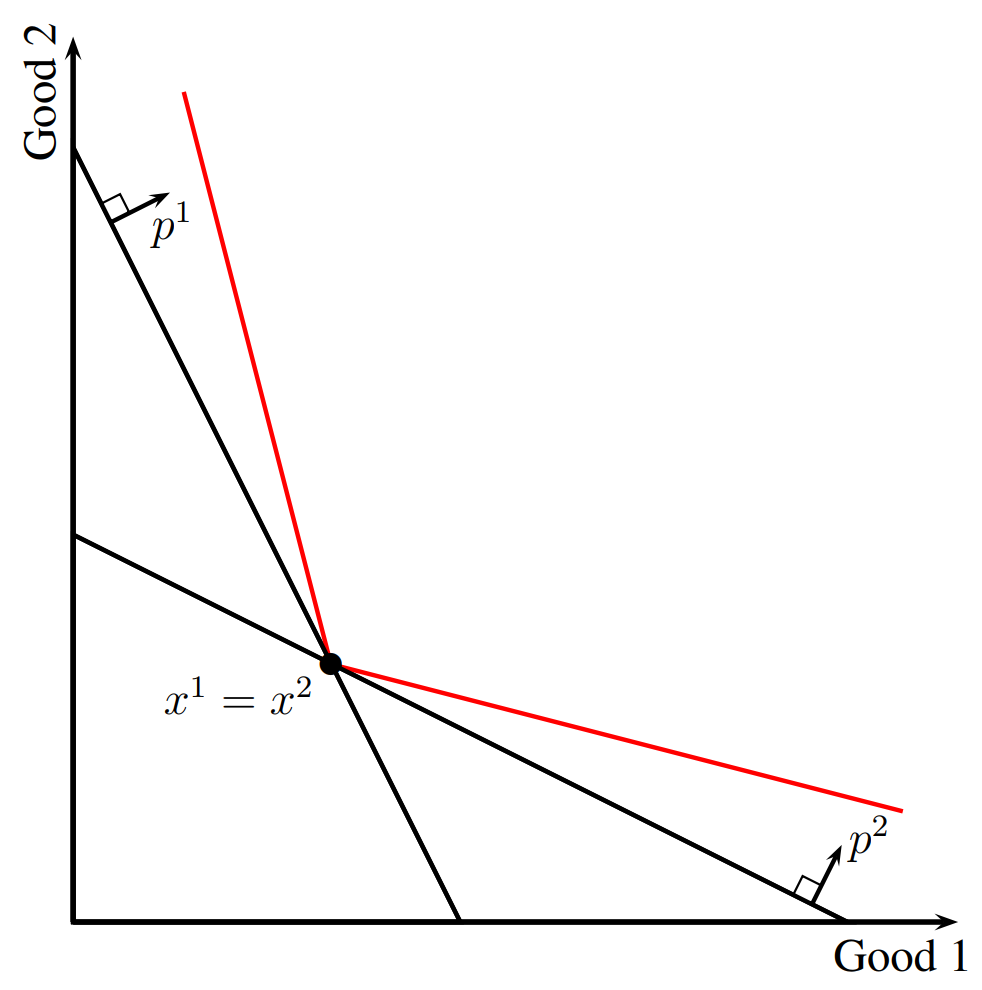}
	\end{center}
	\caption{Non-differentiable rationalization.}
	\label{fig:02}
\end{figure}

Neither GARP nor SARP are sufficient for a dataset to be rationalized by a differentiable utility function. Consider the example depicted in Figure~\ref{fig:02}. Since the consumer selected the same bundle under different prices, the data satisfies both GARP and SARP. However, any utility function rationalizing such a dataset would have an indifference curve with a kink at $\bs{x}^1 = \bs{x}^2$. Thus, it would not be differentiable at this particular point. As shown in \citet{chiappori-rochet87} and \citet{matzkin-richter91}, it is possible to strengthen SARP into a sufficient condition for a differentiable rationalization.
\begin{theorem}[\citealp{chiappori-rochet87} and \citealp{matzkin-richter91}] \label{theorem:cr}
	Suppose that the purchase dataset $D = \big\{(\bs{x}^t, \bs{p}^t)\big\}_{t \leq T}$ satisfies SARP and $\bs{p}^t \neq \bs{p}^s$ implies $\bs{x}^t \neq \bs{x}^s$, for all $t, s$. Then, $D$ can be rationalized by a strictly increasing, strictly concave, and infinitely differentiable utility function $U: \mathbb{R}_{+}^L \to \mathbb{R}$.\footnote{The domain of $U$ is the non-open set $\mathbb{R}_{+}^L$ and so we should be clear about what it means for $U$ to be infinitely differentiable. We say that $U: \mathbb{R}_+^L \rightarrow \mathbb{R}$ is infinitely differentiable if there exists some infinitely differentiable $\ti{U}$ with domain $\mathbb{R}^L$ (instead of $\mathbb{R}_+^L$) so that $U$ is the restriction of $\ti{U}$ to $\mathbb{R}_+^L$.}
\end{theorem}
The rationalizing utility function $U$ supplied in \citet{chiappori-rochet87} is strictly increasing, strictly concave, and infinitely differentiable but its domain is some compact set $K \subseteq \mathbb{R}_{+}^L$ satisfying $\{ \bs{x}^t \}_{t \leq T} \subseteq K$. \citet{matzkin-richter91} in their Theorem $1^{\infty}$ extend the result of \citet{chiappori-rochet87} to what we state in our Theorem \ref{theorem:cr}. 

Theorem \ref{theorem:cr} provides a sufficient but not necessary condition for a dataset $D$ to be rationalized by a differentiable utility function (with some additional properties). In fact, from Theorem $1^{\infty}$ in \citet{matzkin-richter91} it is clear that if the dataset $D = \{ (\bs{x}^t, \bs{p}^t) \}_{t \leq T}$ contains no purchases on the boundary of the consumption space (i.e.\ $\bs{x}^t \gg \bs{0}$ for all $t$) then the condition in Theorem \ref{theorem:cr} is necessary and sufficient. In other words, if it happens that $\bs{x}^t \gg \bs{0}$ for all $t$ then $D$ can be rationalized by a strictly increasing, strictly concave, and infinitely differentiable utility function $U: \mathbb{R}_{+}^L \to \mathbb{R}$ if and only if (i) the data satisfies SARP and (ii) the data satisfies $\bs{p}^t \neq \bs{p}^s$ implies $\bs{x}^t \neq \bs{x}^s$, for all $t, s$. 

See also \cite{Ugarte2023} who finds a necessary and sufficient condition under which a dataset can be rationalized by a differentiable, concave (but not necessarily strictly concave), and strictly increasing utility function.

\subsection{Discrete Consumption} \label{ss:discrete}

From Afriat's Theorem we know that GARP is a necessary and sufficient condition for a dataset to be rationalized by a utility function $U$ with domain $\mathbb{R}_+^L$. This no longer holds when the consumption space is some subset $Y$ of $\mathbb{R}_+^L$. For example, suppose that $Y = \mathbb{N}_+^2$ and let the dataset $D$ consist of observations $\bs{x}^1 = (1,0)$, $\bs{p}^1 = (3,2)$, $\bs{x}^2 = (0,1)$, and $\bs{p}^2 = (2,3)$. Note that in period 1 good 1 is more expensive and yet the consumer buys only this expensive good whereas in period 2 good 2 is more expensive and yet again the consumer only buys the expensive good. This clearly violates GARP (indeed, $\bs{x}^1 \ P \ \bs{x}^2$ and $\bs{x}^2 \ P \ \bs{x}^1$) and yet the data can be rationalized by the strictly increasing utility function $U(x_1, x_2) = x_1 + x_2$.\footnote{To see that $U(x_1,x_2) = x_1 + x_2$ rationalizes the data note first that $U(\bs{x}^1) = U(\bs{x}^2) = 1$ and next note that the affordable bundles in period 1 are $\bs{0}$, $\bs{x}^1$, and $\bs{x}^2$ and so both $\bs{x}^1$ and $\bs{x}^2$ give the largest utility among the affordable bundles and similarly the affordable bundles in period 2 are $\bs{0}$, $\bs{x}^1$, and $\bs{x}^2$ and so again $\bs{x}^1$ and $\bs{x}^2$ give the largest utility among the affordable bundles.} 

This example highlights an important distinction between utility maximization and cost minimization. As $\bs{x}^1 \ R \ \bs{x}^2$ and $\bs{x}^2 \ R \ \bs{x}^1$ in the example it is clear that any rationalizing utility function $U: \mathbb{N}^2 \to \mathbb{R}$ must assign bundles $\bs{x}^1$ and $\bs{x}^2$ the same level of utility. However, this means that in each observation, the consumer is spending more money than necessary, by purchasing the more expensive bundle.  \cite{polisson-quah13} propose an alternative model of consumer choice that induces choices consistent with GARP. 

\begin{proposition}[\citealp{polisson-quah13}]\label{polisson-quah-01}
Let $Y \subseteq \mathbb{R}_+^L$ and let $D = \{ (\bs{x}^t, \bs{p}^t) \}_{t \leq T}$ be a purchase dataset with $\bs{x}^t \in Y$ for all $t$. Suppose there is a function $U: Y \to \mathbb{R}$, a strictly increasing function $v: \mathbb{R}\times \mathbb{R} \to \mathbb{R}$, positive numbers $\{y^t\}_{t \in T}$, and strictly positive numbers $m$ and $\{q^t\}_{t \in T}$ such that, for all $t$, we have
\[
v\big(U(\bs{x}^t), y^t\big) \ = \ \max\Big\{ v\big(U(\bs{x}), y\big) : \bs{p}^t \cdot \bs{x} + q^ty \leq m \Big\}.
\]
Then, the purchase dataset $D$ satisfies GARP.
\end{proposition}
The message of Proposition \ref{polisson-quah-01} is that as long as the consumer maximizes some overall utility that includes a divisible good, the observations of the prices and the demands of the goods in $Y$ obey GARP, even when the space $Y$ is discrete. 

The intuition behind Proposition \ref{polisson-quah-01} is fairly straightforward. Because there is a divisible good (and $v$ is strictly increasing) the consumer cannot be cost-inefficient (i.e.\ in each period they must reach the attained utility level in the cheapest way possible) as any misspent money could be reallocated to the divisible good resulting in an increase in utility. From this observation it is clear that \eqref{eq:rp} holds and so GARP must also hold.

It is worth pointing out that Proposition \ref{polisson-quah-01} imposes no assumptions on the function $U$. In particular, it need not be increasing or concave. Moreover, given that $D$ obeys GARP, one can apply Afriat's Theorem to show that it can be rationalized by a strictly increasing utility function. In fact, as highlighted in the result below, GARP allows us to rationalize the dataset with a utility function that is quasilinear with respect to the (unobserved) divisible good. 

\begin{proposition}[\citealp{polisson-quah13}]\label{polisson-quah-02}
Let $Y \subseteq \mathbb{R}_+^L$ and suppose that the dataset $D$ satisfies GARP. Then, there exists a function $U:Y \to \mathbb{R}$ such that:
\begin{enumerate}
 \item $U$ rationalizes the dataset $D$;
 \item $U$ is consistent with the relations $R$ and $P$ in the sense of \eqref{eq:rp};
 \item $U$ admits a concave and strictly increasing extension to $\mathbb{R}_+^L$;
 \item There are strictly positive numbers $m$ and $\{q^t\}_{t \in T}$ such that, for all $t \in T$,
 \[
	U(\bs{x}^t) + \tfrac{1}{q^t}(m - \bs{p}^t\cdot \bs{x}^t) \ = \ \max\Big\{ U(\bs{x}) + y : \bs{p}^t \cdot \bs{x} + q^ty \leq m \Big\}.
 \]
\end{enumerate}
\end{proposition}

The proof of this result follows almost immediately from Afriat's Theorem. GARP guarantees  that $D$ is rationalizable by a strictly increasing and concave function $U: \mathbb{R}_+^L \to \mathbb{R}$ defined as in \eqref{eq:utility-afriat} over the entire positive orthant $\mathbb{R}^L_+$. Moreover, the function $U$ is consistent with the directly revealed preference relations $R$ and $P$. Clearly, this implies that once we restrict $U$ to $Y \subseteq \mathbb{R}^L_+$, it satisfies conditions 1--3.

To show that condition 4 holds take any $m > \bs{p}^t \cdot \bs{x}^t$, for all $t \in T$, and define $q^t = 1/\lambda_t$, where $\lambda_t$ is the number used in the definition of $U$ in \eqref{eq:utility-afriat}. Notice that, for any $t$, $\bs{x} \in Y$, and $y \geq 0$ that satisfy $\bs{p}^t \cdot \bs{x} + q^t y^t \leq m$, we obtain
\begin{multline*}
U(\bs{x}) + y \ \leq \ U(\bs{x}) + \tfrac{1}{q^t} (m - \bs{p}^t \cdot \bs{x}) \ \leq \ u_t + \lambda_t\bs{p}^t(\bs{x} - \bs{x}^t) + \lambda_t(m - \bs{p}^t \cdot \bs{x}) \\ =  \ u_t + \lambda_t(m - \bs{p}^t\cdot \bs{x}^t) \ = \ u_t +  \tfrac{1}{q^t} (m - \bs{p}^t\cdot \bs{x}^t) \ = \ U(\bs{x}^t) + \tfrac{1}{q^t} (m - \bs{p}^t \cdot \bs{x}^t).
\end{multline*}
and so condition 4 holds.

\subsection{Rationalizing infinite datasets} \label{s:reny}

Up to this point we have only considered datasets $D = \big\{(\bs{x}^t,\bs{p}^t)\big\}_{t \leq T}$ consisting of finitely many observations $T$. Here we relax this restriction and allow for purchase datasets $D = \big\{( \bs{x}^t, \bs{p}^t) \big\}_{t \in {T}}$, where ${T}$ is an arbitrary index set and thus $T$ may contain infinitely many elements. As before, a utility function $U: \eucp \rightarrow \mathbb{R}$ \emph{rationalizes} the purchase dataset $D$ if \eqref{eq:ratio} holds. The revealed preference relations $R$ and $P$ are defined as in Section \ref{s:garp-market} and the definition of GARP is the same as before. \citet{reny15} shows that GARP remains necessary and sufficient for an increasing rationalization even when there are infinitely many observations.\footnote{Recall that a function $U: \mathbb{R}_+^L \rightarrow \mathbb{R}$ is quasiconcave if $U(\alpha \bs{x} + (1-\alpha) \bs{y}) \geq \min( U( \bs{x} ) , U( \bs{y} ) )$ for all $\bs{x}, \bs{y}$.}

\begin{theorem}[\citealp{reny15}] \label{theorem:reny}
	Let $D = \big\{( \bs{x}^t, \bs{p}^t )\big\}_{t \in {T}}$ be an arbitrary (possible infinite) purchase dataset. The following statements are equivalent.
	\begin{enumerate}
		\item $D$ can be rationalized by a locally non-satiated preference relation $\succeq$.

		\item $D$ satisfies GARP.

		\item $D$ can be rationalized by an increasing and quasiconcave utility function $U$.
	\end{enumerate}
\end{theorem}

Note that, although GARP remains necessary and sufficient for a dataset to be rationalized with an increasing utility function, Theorem \ref{theorem:reny}, in contrast to the utility function supplied in Afriat's Theorem, no longer guarantees that the function is continuous, strictly increasing, and concave. 

\citet{reny15} shows through counterexamples that there is little room for strengthening the properties imposed on the rationalizing utility function without imposing further conditions on the data (beyond GARP). In particular, Reny constructs GARP-satisfying datasets $D_1,D_2,D_3$ so that $D_1$ cannot be rationalized by an increasing and lower semicontinuous utility function; $D_2$ cannot be rationalized by an increasing and upper semicontinuous utility function; and $D_3$ cannot be rationalized by an increasing and concave utility function.\footnote{To our knowledge, it is an open question as to whether Reny's result could be strengthened to allow for a quasiconcave and \emph{strictly} increasing rationalization.} Put another way, Afriat's Theorem demonstrates that the hypothesis that utility is continuous, concave, and strictly increasing imposes no additional testable implications on finite purchase datasets beyond what is imposed by the hypothesis that preferences are locally non-satiated. This conclusion is reversed for infinite observation datasets where these properties yield additional testable implications.

Below, we present a slightly modified version of Example 2 from \citet{reny15} which shows that there is a GARP-satisfying purchase dataset which cannot be rationalized by an increasing and upper semicontinuous utility function.\footnote{Recall that a function $U: \eucp \rightarrow \mathbb{R}$ is \emph{upper semicontinuous} if, for all convergent sequences $\bs{x}_n$, we have $\limsup_n U( \bs{x}_n ) \leq U( \lim_n \bs{x}_n )$.}

\addtocounter{example}{1}
\begin{example}\label{example:reny-nousc}
	There are two goods, i.e., $L = 2$. The purchase dataset $D = \big\{(\bs{p}^t, \bs{x}^t) \}_{t=1}^\infty$ consists of elements $\bs{x}^1 = (1,0)$, $\bs{p}^1 = (1,1)$, and $\bs{x}^t = ({3}/{t}, 1-{2}/{t})$, $\bs{p}^t = (1,2)$, for $t \geq 2$. It is easily verified that $\bs{x}^t \ \cancel{R} \ \bs{x}^s$, for all $s > t$ and so the data satisfies GARP. For a contradiction suppose that $D$ is rationalized by an increasing and upper-semicontinuous utility function $U$. As $U$ rationalizes $D$ it must satisfy \eqref{eq:rp} and because $\bs{x}^t \ P \ \bs{x}^s$ for all $t > s$ we see that $U(\bs{x}^1) < U(\bs{x}^2) < U(\bs{x}^3) < \ldots$. Since $\bs{x}^t \rightarrow (0,1)$, upper semicontinuity requires that $U(\bs{x}^1) < \limsup_t U( \bs{x}^t ) \leq U( \lim_t \bs{x}^t ) = U(0,1)$. However, as $\bs{p}^1 \cdot \bs{x}^1 \geq \bs{p}^1 \cdot (0,1)$, it must be that $U(0,1) \leq U(\bs{x}^1)$ and so we conclude $U(\bs{x}^1) < U(0,1) \leq U(\bs{x}^1)$ which is a contradiction. Therefore, $D$ cannot be rationalized by an increasing and upper-semicontinuous utility function. 
\end{example}

\section{Rationalization in general frameworks} \label{s:garp-general}

\subsection{Non-linear prices} \label{ss:fm}

Up to this point prices have been linear. For instance, if one buys the bundle $2 \bs{x}$ it costs twice as much as the bundle $\bs{x}$. Here we consider how to characterize utility maximizing behavior when prices are not necessarily linear. A \emph{price function} is a continuous and strictly increasing function $g: \mathbb{R}_{+}^L \rightarrow \mathbb{R}$ where, intuitively, $g(\bs{x})$ represents the price of (the expenditure associated with) the bundle $\bs{x}$. We consider a \emph{purchase dataset} of the form $D = \{ (\bs{x}^t, g^t) \}_{t \leq T}$ where $\bs{x}^t$ is the consumption bundle purchased by the consumer when faced with the price function $g^t$. 
	
As before, a utility function $U$ rationalizes $D = \{ (\bs{x}^t, g^t) \}_{t \leq T}$ if, for each $t$, the utility derived from $\bs{x}^t$ exceeds the utility of any other affordable bundle, i.e.,\, $U(\bs{x}^t) \geq U(\bs{x})$ for all $\bs{x}$ satisfying $g^t(\bs{x}^t) \geq g^t(\bs{x})$. Similarly, a preference relation $\succeq$ rationalizes $D$ if, for all $t$, the bundle chosen $\bs{x}^t$ is preferred to every other affordable bundle (i.e.\ $\bs{x}^t \succeq \bs{x}$ for all $\bs{x}$ satisfying $g^t(\bs{x}^t) \geq g^t(\bs{x})$). Note that if all price functions are positive linear functions (i.e.\ $g^t(\bs{x}) = \bs{p}^t \cdot \bs{x}$ for some price vector $\bs{p}^t \in \mathbb{R}_{++}^L$) then our dataset $D$ is as in Section \ref{s:garp-market} and thus Afriat's Theorem applies. In fact, even when $g^t$ are not linear a version of GARP still characterizes strictly increasing utility maximization.

We define $R$ and $P$ in essentially the same way as in Section \ref{s:garp-market}. To be precise, $\bs{x}^t \ R \ \bs{y}$ means $\bs{y}$ was affordable when $\bs{x}^t$ was purchased (i.e.\ $g^t(\bs{x}^t) \geq g^t(\bs{y})$) and $\bs{x}^t \ P \ \bs{y}$ means $\bs{y}$ was cheaper than $\bs{x}^t$ in period $t$ (i.e.\ $g^t(\bs{x}^t) > g^t(\bs{y})$). It is easy to see that \eqref{eq:rp} holds for all strictly increasing and rationalizing utility functions $U$ and thus if $D$ can be rationalized by a strictly increasing utility function then $D$ satisfies GARP (where GARP continues to be as in Definition \ref{def:GARP1} only now using our generalized definitions of $R$ and $P$). As shown in \citet{forges09} the reverse implication also holds.\footnote{The functions $g^t$ in \citet{forges09} are normalized so that $g^t(\bs{x}^t) = 0$. This normalization has no impact on the result in Theorem \ref{theorem:forges-minelli} and can easily be imposed if so desired by defining $\ti{g}^t$ by $\ti{g}^t(\bs{x}) = g^t(\bs{x}) - g^t(\bs{x}^t)$ and working with the normalized dataset $\{ (\bs{x}^t, \ti{g}^t) \}_{t \leq T}$.}\textsuperscript{,}\footnote{The functions $g^t$ do not have to represent prices. Instead, $g^t$ can be thought of as any function which describes the constraint set faced by the consumer in period $t$ in the sense that in each period $t$ the consumer must select a bundle in the constraint set $\{ \bs{x} \in \mathbb{R}_{+}^L: g^t(\bs{x}^t) \geq  g^t(\bs{x}) \}$. \citet{forges09} provide some analysis of the types of constraint sets which can be described in this fashion.}

\begin{theorem}[\citealp{forges09}] \label{theorem:forges-minelli}
	For any dataset $D = \{ (\bs{x}^t, g^t) \}_{t \leq T}$ the following statements are equivalent.
	\begin{enumerate}
		\item The dataset $D$ can be rationalized by a locally non-satiated preference relation.
		\item The dataset $D$ satisfies GARP.
		\item There are numbers $\{u^t\}_{t \leq T}$ and strictly positive numbers $\{\lambda^t\}_{t \leq T}$ such that
		\begin{equation*} \label{eq:afriat-ineqs3}
			u^s \ \leq \ u^t + \lambda^t (g^t( \bs{x}^s) - g^t(\bs{x}^t )), \qquad \text{ for all } s,t \in \{1,2,\ldots, T\}.
		\end{equation*}
		\item The dataset $D$ can be rationalized by a continuous and strictly increasing utility function. 
	\end{enumerate}
\end{theorem}
Implication 4 $\Rightarrow 1$ is obvious. Implication 1 $\Rightarrow$ 2 and implication 2 $\Rightarrow$ 3 can be shown using a similar approach to what we employed in our proof of Afriat's Theorem (just replace $\bs{p}^j \cdot ( \bs{x}^k - \bs{x}^j )$ and $\bs{p}^i \cdot (\bs{x}^j - \bs{x}^i)$ with $g^j( \bs{x}^k ) - g^j(\bs{x}^j)$ and $g^i( \bs{x}^j ) - g^j(\bs{x}^i)$, respectively in equations \eqref{eq:u-def2} and \eqref{eq:lambda-def2}). Finally, given the numbers $u^t, \lambda^t$, it is easy to prove that the function 
\begin{equation} \label{eq:fm-utility}
	U(\bs{x}) := \min_{t}\big\{u^t + \lambda^t ( g^t(\bs{x}) - g^t(\bs{x}^t) ) \big\}
\end{equation}
is strictly increasing, continuous, and rationalizes the data.

The utility function supplied in statement 4 of Theorem \ref{theorem:forges-minelli} is not concave whereas the utility function in statement 4 of Afriat's Theorem is concave. Indeed, when prices are non-linear concave and strictly increasing utility has additional testable restrictions beyond GARP. However, if the price functions $g^t$ are concave then the utility function $U$ constructed in \eqref{eq:fm-utility} is also concave and thus GARP is necessary and sufficient for a \emph{concave}, strictly increasing, and continuous rationalization when each of the price functions also satisfy these properties. A version of this observation was presented originally in \citet{matzkin91} in a setup with differentiable rationalizations (as in Section \ref{ss:sarp}). See also \cite{BelgiansConvex} for a study of the testable implications of concave rationalizations.

Theorem \ref{theorem:forges-minelli} is largely unchanged even when the dataset $D$ has infinitely many observations. In particular, suppose we have a dataset $D = \{ (\bs{x}^t, g^t) \}_{t \in T}$ where $T$ is an arbitrary index set and $g^t$ are prices functions. In this context \citet{reny15} shows that GARP is necessary and sufficient for there to exist an increasing utility function which rationalizes the data. 

\subsection{Generalized Monotonicity and Abstract Choice Spaces} \label{ss:noq}

Consider an arbitrary consumption space $X$ and suppose the researcher observes a \emph{choice dataset} $D = \big\{( \bs{x}^t, B^t )\big\}_{t \in T}$, where $T$ is an arbitrary index set and $\bs{x}^t \in X$ denotes an alternative selected from the constraint set $B^t \subseteq X$ (and so, we require $\bs{x}^t \in B^t$ for all $t \in T$). The choice dataset $D$ is \emph{rationalized} by utility function $U: X \to \mathbb{R}$ if, for any $t \in T$, the bundle chosen $\bs{x}^t$ yields more utility than any other bundle which could have been chosen (i.e.\ $U(\bs{x}^t) \geq U(\bs{x})$ for all $\bs{x} \in B^t$). Alternatively, the dataset is \emph{rationalized} by the preference relation $\succeq$ if the bundle selected $\bs{x}^t$ is preferred to every other bundle which could have been chosen (i.e.\ $\bs{x}^t \succeq \bs{x}$ for all $\bs{x} \in B^t$). When $X = \mathbb{R}_+^L$ and each constraint set $B^t$ takes the form $\{ \bs{x} \in \mathbb{R}_{+}^L: \bs{p}^t \cdot \bs{x}^t \geq \bs{p}^t \cdot \bs{x} \}$ for some price vector $\bs{p}^t \in \mathbb{R}_{++}^L$ then our new notion of rationalization coincides with the old one.

As in the preceding sections, unless we impose some additional restrictions, any dataset is rationalizable with a \emph{constant} utility $U$. For this reason we restrict our attention to utilities that are strictly increasing with respect to some ordering $\Tre$ on $X$. For example, within Euclidean spaces it may be sensible to consider preferences in which ``more is better'', i.e., we have $\Tre = \ \geq$. When studying choice under risk, the natural order is that of first order stochastic dominance over lotteries. When studying choice over time, one could consider the order of impatience (specified below in Example \ref{example:impatience}). Here we present general results that are applicable to a great variety of such orderings.

Formally, let $\Tre$ be a preorder on $X$, i.e., a reflexive and transitive binary relation, with its asymmetric part denoted by $\Tr$.\footnote{So, $\bs{x} \Tr \bs{y}$ means $\bs{x} \Tre \bs{y}$ and $\bs{y} \ \cancel{\Tre} \ \bs{x}$.} A utility function $U: X \rightarrow \mathbb{R}$ is \emph{strictly $\Tre$-increasing} if $\bs{x} \Tre \bs{y}$ implies $U(\bs{x}) \geq U(\bs{y})$, and $\bs{x} \Tr \bs{y}$ implies $U(\bs{x}) > U(\bs{y})$. Similarly, a preference relation $\succeq$ over $X$ is \emph{strictly $\Tre$-increasing} if $\bs{x} \Tre \bs{y}$ implies $\bs{x} \succeq \bs{y}$, and $\bs{x} \Tr \bs{y}$ implies $\bs{x} \succ \bs{y}$.\footnote{Note that the concept of ``strictly $\geq$-increasing'' (where $\geq$ is the usual Euclidean order on $\eucp$) is the same as what we simply refer to as ``strictly increasing."} Below we present two examples of relevant economic setups.

\begin{example} \label{example:fosd}
	Suppose there are $L$ states of the world and let $\bs{x} = (x_1,x_2,\ldots, x_L) \in \eucp$ represent a state contingent consumption bundle where $x_{\ell} \in \mathbb{R}_+$ denotes the amount of consumption (or money) received in state $\ell$. Let $\bs{\pi} = (\pi_1, \pi_2,\ldots,\pi_L) \in \mathbb{R}_{+}^L$ be a probability vector (i.e.\ $\sum_{\ell=1}^L \pi_{\ell} = 1$) which represents the likelihood of the different states of the world occurring. A state{-}contingent consumption bundle $\bs{x}$ \emph{first order stochastically dominates} $\bs{y}$ if, for each number $\alpha \in \mathbb{R}$ the probability that $\bs{x}$ pays out more than $\alpha$ is greater than the probability that $\bs{y}$ pays out more than $\alpha$. Let us write it as $\bs{x} \Tre_{FOSD} \bs{y}$, with its strict part denoted by $\bs{x} \Tr_{FOSD} \bs{y}$. It is natural (and required by many models of consumer choice under risk) to think that a consumer should prefer $\bs{x}$ to $\bs{y}$ whenever $\bs{x} \Tre_{FOSD} \bs{x}'$. In other words, the utility function of a consumer ought to be $\Tre_{FOSD}${-}strictly increasing.
\end{example}

\begin{example} \label{example:impatience}
	Suppose there are two time periods $t = 0, 1$ and let $\bs{x} = (x_0,x_1) \in \mathbb{R}_+^2$ denote a stream of consumption where $x_{0}$ is the amount of consumption (or money) received now and $x_1$ is the amount of consumption received later. It is natural to think that an impatient consumer should prefer the stream $(M, m)$ to $(m,M)$, whenever $M \geq m$. This preference for receiving larger amounts earlier can be summarized with a preorder. Specifically, let $\Tre_{imp}$ denote the smallest preorder which extends the natural partial order $\geq$ over $\mathbb{R}^2$, and satisfies $(M,m) \Tre_{imp} (m,M)$, for all  $M \geq m$.\footnote{We say that a binary relation $\succeq$ extends  a second binary relation $\succeq'$ if $x \succeq' x'$ implies $x \succeq x'$. The existence of the preorder $\Tre_{imp}$ is easy to establish. Let $\Tre_{imp}^*$ be the preorder defined by $(M,m) \Tre_{imp}^* (m,M)$, for any $M \geq m$. It is easy to see that $\Tre_{imp}$ is the transitive closure of $\geq \cup \ \Tre_{imp}^*$.}
	%
	
\end{example}

As in the preceding sections, it is convenient to discuss the testable restrictions of utility maximization by first introducing some revealed preference relations. Let $X$ be an arbitrary space endowed with a preorder $\Tre$ and let $D = \big\{( \bs{x}^t, B^t )\big\}_{t \in T}$ be a (possibly infinite) choice dataset. We write $\bs{x}^t \ R_{\Tre} \ \bs{y}$ if there exists $\bs{z} \in B^t$ such that $\bs{z} \Tre \bs{y}$, in other words, when $\bs{x}^t$ is chosen over an alternative $\bs{z}$ that dominates $\bs{y}$ with respect to the ordering $\Tre$. Note that if $\bs{y}\in B^t$ then $\bs{x}^t \ R_{\Tre} \ \bs{y}$, but our definition of $R_{\Tre}$ allows for $\bs{y}$ \emph{not} to belong to $B^t$. Similarly, we write $\bs{x}^t  \ P_{\Tre} \ \bs{y}$ if there exists $\bs{z} \in B^t$ such that $\bs{z} \Tr \bs{y}$, in other words, when $\bs{x}$ is chosen over an alternative $\bs{z}$ that \emph{strictly} dominates $\bs{y}$ with respect to $\Tre$.\footnote{Notice that, within the original setup of \cite{afriat67} (as in Section \ref{ss:afriat}) or \cite{forges09} (as in Section \ref{ss:fm}), the revealed preference relations $R_{\geq}, P_{\geq}$ coincide exactly with $R, P$, defined in the corresponding sections.}

It is straightforward to show that \eqref{eq:rp} holds (replacing $R$ and $P$ with $R_{\Tre}$ and $P_{\Tre}$) provided $U$ is strictly $\Tre$-increasing and rationalizes the data. Indeed, suppose that $\bs{x}^t \ R_{\Tre} \ \bs{y}$. By definition of the relation, there is $\bs{z} \in B^t$ so that  $\bs{z} \Tre \bs{y}$. Since $U$ rationalizes $D$ and is $\Tre$-strictly increasing, we have $U(\bs{x}^t) \geq U(\bs{z}) \geq U(\bs{y})$ and so $U(\bs{x}^t) \geq U(\bs{z})$ as claimed. A similar argument establishes that $\bs{x}^t \ P_{\Tre} \ \bs{y}$ implies $U(\bs{x}^t) > U(\bs{y})$. From these observations we see that the following acyclicity condition holds whenever the data can be rationalized by a strictly $\Tre$-increasing utility function.

\begin{definition}[$\Tre${-}GARP] \label{def:garp-Tre}
	A dataset $D = \big\{( \bs{x}^t, B^t )\big\}_{t \in T}$ satisfies \emph{the generalized axiom of revealed preferences} for $\Tre$ (henceforth, $\Tre${-}GARP) if there is \emph{no} sequence $t_1,t_2,\ldots, t_K$ in $T$ such that
	\begin{equation} \label{eq:garp-Tre}
		\bs{x}^{t_1} \ R_{\Tre} \ \bs{x}^{t_2} \ R_{\Tre} \ \bs{x}^{t_3} \ R_{\Tre} \ \ldots \ R_{\Tre} \ \bs{x}^{t_K}, \text{ and } \ \bs{x}^{t_K} {P}_{\Tre} \ \bs{x}^{t_1}.
	\end{equation}
\end{definition}

Note that, as in the original formulation of GARP, the above condition excludes particular cycles induced by the revealed preference relations. In fact, within the setup of \cite{afriat67} (as in Section \ref{ss:afriat}) or \cite{forges09} (as in Section \ref{ss:fm}), $\Tre${-}GARP coincide exactly with GARP, whenever we set $\Tre = \ \geq$.

It is clear that $\Tre${-}GARP is necessary for a dataset to be rationalizable with a $\Tre${-}strictly increasing preference relation. As shown in \citet{nishimura-oK-quah17} it happens to also be sufficient.

\begin{theorem}\label{thm:NOQ1}
	Let $X$ be an arbitrary space endowed with a preorder $\Tre$ and let $D = \big\{( \bs{x}^t, B^t )\big\}_{t \in T}$ be a choice dataset. The data $D$ can be rationalized by a strictly $\Tre${-}increasing preference relation $\succeq$ if, and only if, it satisfies $\Tre${-}GARP.
\end{theorem}

Necessity of $\Tre${-}GARP for rationalization follows from our previous discussion. The proof of sufficiency involves applying Szpilrajn's extension theorem to the transitive closure of $\Tre \ \cup \ R_{\Tre}$. While Theorem \ref{thm:NOQ1} shows that $\Tre$-GARP is enough to find a strictly $\Tre$-increasing preference relation which rationalizes the data, the theorem does not claim that this preference relation can be represented by a utility function. In order to guarantee this, more structure must be imposed on the data and the choice space. The following is from \citet{nishimura-oK-quah17}. 

\begin{theorem}\label{thm:NOQ2}
	Let $X$ be a separable and locally compact metric space and let $\Tre$ be a continuous preorder.\footnote{ %
		This is to say that $\Tre$ is a closed subset of $X \times X$.
	}
	Moreover, suppose that the dataset  $D = \big\{( \bs{x}^t, B^t )\big\}_{t \in T}$ has finitely many observations (i.e.\ the index set $T$ is finite) and the set $B^t$ is compact, for all $t \in T$. Then, the dataset $D$ can be rationalized by a continuous and strictly $\Tre$-increasing utility function $U:X \to \mathbb{R}$ if and only if $D$ satisfies $\Tre${-}GARP.
\end{theorem}

Importantly, $\mathbb{R}_{+}^L$ is a separable and locally compact metric space and thus Theorem \ref{thm:NOQ2} can be used to recover part of Afriat's Theorem. Note however that the consumption space $X$ in this theorem is not a vector space and so, unlike Afriat's Theorem, it has nothing to say about the concavity of the rationalizing utility function.  

The proof that $\Tre$-GARP is sufficient for a dataset to be rationalized by a continuous and $\Tre$-strictly increasing utility can be carried out by applying Levin's Theorem.  This theorem  states that any continuous preorder admits a complete extension that can be represented with a continuous utility function.  The proof of Theorem \ref{thm:NOQ2} proceeds by first showing that the transitive closure of $R \cup \Tre$ is a continuous preorder and then completing this preorder by applying Levin's Theorem. The final step involves checking that this preorder (equivalently, the continuous utility function that is its representation) rationalizes the data.  

The theorem's feature of allowing the rationalizing utility function $U$ to be increasing in an order of interest $\Tre$ has proven to be useful in applied studies. \citet{polisson-quah-renou20} use Theorem \ref{thm:NOQ2} to test for utility maximization over lotteries where the utility function is required to be increasing in the ordering of first order stochastic dominance and \citet{lanier22} use Theorem \ref{thm:NOQ2} to test for utility maximization over risky temporal streams of consumption where various orders of interest are imposed on the rationalizing utility function.

We return to Example \ref{example:fosd} to further discuss Theorem \ref{thm:NOQ2}. 

\addtocounter{example}{-2}
\begin{example}[continued]
	Suppose there are two equally likely states of the world, i.e., $\bs{\pi} = (\tfrac{1}{2}, \tfrac{1}{2})$, and suppose we observe a consumer purchase the state-contingent consumption bundle $(1,0)$ under prices $(2,1)$. Therefore, $D$ consist of a single observation $\bs{x}^1 =  (1,0), \bs{p}^1 = (2,1)$. In particular, the set $D$ trivially satisfies GARP and, thus, Theorem \ref{theorem:afriat} guarantees that there is a continuous, concave, and strictly increasing utility function which rationalizes the dataset.
	
	However, one can easily check, that any such utility would not be consistent with the first order stochastic dominance order. Indeed, note that the bundle $(0,2)$ was affordable when $(1,0)$ was purchased under the prevailing prices. Therefore, any utility $U$ rationalizing the choice must satisfy $U(1,0) \geq U(0,2)$. Given that $(0,2) \Tr_{FOSD} (1,0)$, any such utility must violate first order stochastic dominance order. Therefore, the original formulation of GARP is not sufficient for this stricter form of rationalization. In contrast, $\Tre_{FOSD}${-}{GARP} provides the exact characterization of this model and this property is violated in this example, since $(1,0)\, P_{\Tre_{FOSD}}\, (1,0)$. 
\end{example}


\section{Related results and extensions}\label{s:extension}

In the previous sections we have shown that the essence of utility maximization and its testable implications are summarized with an acyclicity condition --- the generalized axiom of revealed preference and its variations. In this section we consider several models that are, in one way or another, somewhat removed from the standard setup and show that GARP-like properties continue to play a role in their characterization.

\subsection{Measuring departures from rationality} \label{ss:ccei}

It is very common in empirical studies for purchase datasets to violate GARP and, thus, are not (exactly) consistent with utility maximization.\footnote{See Chapter 5 in \cite{EcheniqueChambersBook} for a general discussion on the empirical analysis. More recently, inconsistencies with utility maximization have been reported in \cite{HalevyPersitzZrill2017}, \cite{EcheniqueImaiSaitoExperiment},  \cite{FeldmanRehbeck}, \cite{Zrill2020},  \cite{DemboKarivPolissonQuah2021}, and \cite{Cappelen2021}.}  Perhaps this is not entirely surprising, given the deterministic nature of GARP, which is either satisfied by a dataset or not.  When faced with a violation of GARP, one may be interested in measuring the severity of the dataset's inconsistency with utility-maximization.  Part of the revealed preference literature has been devoted to formulating and answering this question, which is obviously important for empirical applications.\footnote{ See \cite{Afriat1973},  \cite{HoutmanMaks1985}, \cite{Varian1990}, \cite{MP2011}, \cite{DeanMartinMCI}, \cite{ApesteguiaBallester2015}, \cite{EcheniqueSaitoEU, EcheniqueSaitoEDU}, \cite{Dziewulski_CCEI}, \cite{AllenRehbeckQL, AllenRehbeckCCEI},  \cite{ClippelRozen2020}, or \cite{Ribeiro-comparative}.}


Arguably, the most common measure of departures from rationality is the \emph{critical cost-efficiency index} (CCEI, also known as \emph{Afriat's efficiency index}), introduced in \cite{Afriat1973} to evaluate violations of utility maximization within the standard consumer demand framework.\footnote{Among others, CCEI was employed in  \cite{Sippel1997}, \cite{GARPkids}, \cite{AndreoniMiller2002}, \cite{Choi2007}, \cite{Fisman2007}, \cite{Ahn2014}, \cite{Choi2014}, \cite{Belgians2017}, \cite{EcheniqueImaiSaitoExperiment}, \cite{Belgians2020}, \cite{DemboKarivPolissonQuah2021}, and \cite{Cappelen2021}.}
Consider the setup discussed in Section \ref{s:garp-market}, where a purchase dataset is given by  $D = \big\{( \bs{x}^t, \bs{p}^t )\big\}_{t \leq T}$, with $\bs{x}^t \in \mathbb{R}_+^L$ denoting a consumption bundle selected by the consumer under the prevailing prices $\bs{p}^t \in \mathbb{R}_{++}^L$. Take any number $e \in [0,1]$ -- an \emph{efficiency parameter} -- and define the following binary relation:\, $\bs{x}^t \   R_e \ \bs{y}$, whenever $\bs{p}^t\cdot \bs{y} \leq e( \bs{p}^t\cdot \bs{x}^t )$, i.e., the bundle $\bs{x}^t$ dominates $\bs{y}$ in this sense if the former was chosen when the value of the latter was at most $e\cdot 100\%$ of the value of $\bs{x}$, under the prevailing prices.  Analogously, we denote $\bs{x}^t \  P_e \ \bs{y}$, whenever $\bs{p}^t \cdot \bs{y} < e(\bs{p}^t\cdot \bs{x}^t)$. Clearly, whenever $e = 1$, the above relations coincide with those defined in Section \ref{ss:afriat}.


Given the relations $R_e, P_e$, one can easily define an acyclicity condition analogous to GARP. The dataset $D$ satisfies GARP for the efficiency parameter $e$ ($e${-}GARP), if there is \emph{no} sequence $t_1,t_2,\ldots, t_K$ such that $\bs{x}^{t_1} \ R_e \ \bs{x}^{t_2} \  R_e \ \bs{x}^{t_3} \ R_e \  \ldots \ R_e \ \bs{x}^{t_K}$, and $\bs{x}^{t_K} \ {P}_e \  \bs{x}^{t_1}$; in other words, the relations $R_e$ and $P_e$ are required to obey the property highlighted in Definition \ref{def:GARP1}.  Clearly, $e${-}GARP is weaker than GARP in the sense that, the latter holds only if so does the former and, more generally, a dataset satisfies $e'${-}GARP if it obeys $e${-}GARP for any $e'\leq e$. This is clear because $R_{e'}\subseteq R_e$ and $P_{e'}\subseteq P_e$.  

Clearly, $e${-}GARP is not sufficient for a dataset to be rationalizable with utility maximization. However, it characterizes a weaker notion of rationality.

\begin{proposition} \label{prop:e-garp}
Take any efficiency index $e \in (0,1]$. A dataset $D$ satisfies $e${-}{GARP} if, and only if, if there is a strictly increasing utility $U:\mathbb{R}_+^{L} \to \mathbb{R}$ that {\em $e$-rationalizes $D$}, by which we mean the following:\, for all $t$,
\[
e(\bs{p}^t\cdot \bs{x}^t) \geq \bs{p}^t\cdot \bs{y} \ \text{ implies } \ U(\bs{x}^t) \geq U(\bs{y}).
\]
\end{proposition}

Proofs of this proposition appear in \cite{Afriat1973}, \citet{HalevyPersitzZrill2017}, \citet{polisson-quah-renou20}, and \citet{lanier22}.  This result says that $e${-}GARP is necessary and sufficient for the data to be partially rationalized in the sense that the observed choice $\bs{x}^t$ gives higher utility than any bundle that costs less than $e \cdot 100\%$ of its value. This rationalization is only partial since it leaves open the possibility that, at some observation $t$, a bundle valued between $\bs{p}^t\cdot \bs{x}^t$ and $e(\bs{p}^t\cdot \bs{x}^t)$ gives higher utility than $\bs{x}^t$.\footnote{This imperfect utility-maximization could also be interpreted as imperfect cost-minimization subject to a utility target (see \cite{PolissonQuahCCEI}).} The CCEI is defined as the supremum over all efficiency indices $e \in (0,1]$ for which the dataset admits $e${-}rationalization by a strictly increasing utility function.   



\cite{Dziewulski_CCEI} shows that $e$-rationalizations can be understood in terms of \emph{just-noticeable differences}, in a model where consumers have limited ability to distinguish similar alternatives.  In \citet{cherchye23}, the CCEI is used in a statistical test of the utility maximization hypothesis. The notion of CCEI can be modified to measure the severity of a dataset's departure from rationalization by utility functions satisfying specific properties.   For example, we may be interested in the greatest $e$ at which $D$ can be $e$-rationalized by a utility function that is strictly increasing with respect to some given preorder $\Tre$ (as in Section \ref{ss:noq}); for applications of such an extension of the CCEI and a guide to its computation, see \citet{polisson-quah-renou20} and \citet{lanier22}.

\subsection{Acyclic strict preference relation} \label{ss:acyclic-strict-prefs}

In some instances, one may consider the `weak' revealed preference relation $R$ to be less informative of the agent's true preferences than the strict preference $P$. This may be due to a level of imprecision when stating indifferences or the subject's inability to perfectly discriminate among similar  alternatives. As a result, one may want to investigate implications of an acyclicity condition that is imposed solely on the strict revealed relation $P$.

This is addressed in \cite{AUM-main}. Formally, the condition requires that there is \emph{no} sequence $t_1,t_2,\ldots, t_K$ such that $\bs{x}^{t_1} \ P \ \bs{x}^{t_2} \ P \ \bs{x}^{t_3} \ P \  \ldots \ P \ \bs{x}^{t_K} \ P \ \bs{x}^{t_1}$.
The restriction implicitly assumes that only the revealed strict comparisons convey reliable information about the preferences of the individual, while the weak ones may be subject to imprecision, vagueness of judgement, or incommensurability.  Therefore, only $P$ is required to exhibit some form of consistency. \cite{AUM-main} shows that acyclicity of $P$ is equivalent to the dataset being rationalizable with \emph{approximate utility maximization}. That is, there is a utility $U$ and a positive threshold function $\delta$ such that (i) $\bs{x} > \bs{y}$ implies $U(\bs{x}) > U(\bs{y}) + \delta(\bs{y})$ and (ii) $\bs{p}^t \cdot \bs{x}^t \geq \bs{p}^t \cdot \bs{y}$ implies $U(\bs{x}^t) + \delta(\bs{y})  \geq  U(\bs{y})$, for all $t$. The alternative $\bs{x}^t$ is chosen only if its utility is at most $\delta(\bs{y})$ utils lower than that of any other affordable option $\bs{y}$. This representation appeals to the idea of imperfect discrimination, suggesting that the individual discerns between two alternatives only if they yield a sufficiently different utility.\footnote{ %
 The reader may recognize that this model is following the interval order representation of preferences proposed in \cite{Fishburn1970}.}

\subsection{Revealed price preference} \label{ss:rpp}

So far we have been focusing on how individuals reveal their preferences over consumption bundles.  \cite{deb-kitamura-quah-stoye22} explore the implications of an agent who has a well-behaved revealed preference {\em over prices}. Suppose that we observe a consumer choosing bundle $\bs{x}^t$ at prices $\bs{p}^t$,  and $\bs{x}^s$ at prices $\bs{p}^s$. The price $\bs{p}^s$ is revealed preferred to $\bs{p}^t$, denoted by $\bs{p}^s \ R_p \ \bs{p}^t$, whenever $\bs{p}^s \cdot \bs{x}^t \leq \bs{p}^t \cdot \bs{x}^t$; in other words, $\bs{p}^s$ is revealed prefered to $\bs{p}^t$ whenever the cost of purchasing bundle $\bs{x}^t$ is lower at $\bs{p}^s$ than at $\bs{p}^t$ (the prevailing price at observation $t$). The revealed preference relation is said to be strict, and denoted by $\bs{p}^s \ P_p \ \bs{p}^t$, whenever the cost is strictly lower.

Given the revealed preference relations $R_p$, $R_p$, one may be interested in exploring the implications of imposing a consistent condition on these relations. In particular, suppose $R_p$, $P_p$ are free of cylces, in the sense defined in Section \ref{ss:afriat}, i.e., there is \emph{no} sequence $t_1,t_2,\ldots, t_K$ such that $\bs{p}^{t_1} \ R_p \ \bs{p}^{t_2} \ R_p \ \bs{p}^{t_3} \ R_p \  \ldots \ R_p \ \bs{p}^{t_K}$ and $\bs{p}^{t_K} \ P_p \  \bs{p}^{t_1}$. \cite{deb-kitamura-quah-stoye22} show that this condition, dubbed the \emph{generalized axiom of price preference} (or GAPP), is necessary and sufficient for the dataset to be rationalized with an \emph{expenditure-augmented utility function}; formally, there is a strictly increasing utility $U:\mathbb{R}_+^L \times \mathbb{R}_- \to \mathbb{R}$ such that $U(\bs{x}^t, -\bs{p}^t\cdot \bs{x}^t) \geq U(\bs{y}, -\bs{p}^t\cdot \bs{y})$, for all $\bs{y} \in \mathbb{R}_+^L$ and $t$. Notice that in this notion of rationalization, the consumer does not have a budget constraint and could (in principle) spend as much as he likes, but he does not spend an infinite amount because higher expenditure leads to dis-utility since the utility function has expenditure as its final argument and is strictly decreasing in expenditure. The expenditure-augmented utility function could be thought of as generalization of the familiar quasilinear form $U(\bs{x},-\bs{p}\cdot \bs{x})=V(\bs{x})-\bs{p}\cdot \bs{x}$, for some increasing function $V$.

\subsection{GARP in mechanism design} \label{ss:mech}

Applications of the revealed preference theorems surveyed in this paper go beyond consumer/decision theory. \cite{DebMishra2014} apply this approach to study mechanism design.   

Suppose an agent faces a mechanism designer\footnote{\citet{DebMishra2014} consider multiple agents each with their own type-space and collection of utility functions. We consider a single agent because this significantly simplifies the presentation.} who is unaware of the agent's type $t \in T$. An agent of type $t \in T$ has utility function $v^{t}: A \rightarrow \mathbb{R}$ where $A$ is some set of alternatives. A \emph{social choice function} (SCF) $f: T \rightarrow A$ maps types to alternatives. The output of the SCF $f(t)$ is the alternative which the designer would like to select if she knew that the agent was type $t$. To improve readability we write $f^t$ instead of $f(t)$.

A key assumption in \citet{DebMishra2014} is that, while the agent's type is not observable, the payoff received by the agent from the alternative selected is observable and thus can be contracted on. A \emph{contingent contract} is a collection $\{ c^t \}_{t \in T}$ where $c^t: \mathbb{R} \rightarrow \mathbb{R}$ is strictly increasing for each $t$. When the agent reports $t \in T$ the contingent contract $\{ c^t \}_{t \in T}$ gives the agent with type $s \in T$ a value of $c^t ( v^{s}( f^t ) )$. In other words, the contingent contract rewards the agent according to (i) the report of the agent and (ii) the payoff that the agent actually derives from $f^t$.\footnote{It is true that a mechanism designer may surmise from the observed payoff that the agent has not reported his type truthfully; the assumption is that the contract must still reward the agent in a way that is strictly increasing in the agent's payoff.  For a discussion of this issue see \citet{DebMishra2014}.}  The contingent contract $\{c^t\}_{t \in T}$ is a \emph{linear contract} if there are numbers $\{ u^t \}_{t \in T}$ and $\{ \lambda^t \}_{t \in T}$ where $\lambda^t \in (0,1]$ and $c^t( v) = u^t + \lambda^t v$ for all $v \in \mathbb{R}$ and $t \in T$.

We collect the utility functions $v^t$ and the social choice function $f^t$ into a dataset $D = \{ (v^t, f^t) \}_{t \in T}$. We say that $D$ is \emph{implemented} by the contingent contract $\{ c^t \}_{t \in T}$ if, for all $s,t \in T$ the utility derived by the agent from reporting his true type is weakly larger than the utility he derives from any other report. That is, $D$ is implemented by $\{ c^t \}_{t \in T}$ if $c^t \big( v^{t}(  f^t ) \big) \geq c^s \big( v^{t}( f^s ) \big)$ for all $s,t$. Note that if the contingent contract is linear then this condition becomes
\begin{equation} \label{eq:implementable}
	u^t + \lambda^t v^t(f^t) \geq u^s + \lambda^s v^t(f^s)
\end{equation}
which has a similar structure to the Afriat inequalities of \eqref{eq:afriat-ineqs}. 

It turns out that $D = \{ (v^t, f^t) \}_{t \in T}$ can be implemented if and only if it satisfies a GARP-like acyclicity condition. We write $v^t \ R \ v^s$ to mean $v^t( f^s ) \geq v^s( f^s )$ and further we write $v^t \ P \ v^s$ to mean $v^t( f^s ) > v^s( f^s )$. We say that $D = \{( v^t, f^t) \}_{t \in T}$ is \emph{acyclic} if the relations $R$ and $P$ satisfy the conditions in Definition \ref{def:GARP1} (i.e. there is no sequence $t_1,t_2,\ldots, t_K \in T$ so that $v^{t_1} \ R \ v^{t_2} \ R \ \ldots \ R \ v^{t_K}$, and $v^{t_K} \ P \ v^{t_1}$). The following is a version of Theorem 1 in \citet{DebMishra2014}.\footnote{Again we remark that \citet{DebMishra2014} allow for multiple agents and so their theorem is significantly more involved.}
\begin{theorem} \label{theorem:deb-mishra}
	$D = \{ (v^t, f^t) \}_{t \in T}$ (for finite $T$) can implemented by a contingent contract if and only if it is acyclic. Moreover, if $D$ can be implemented by a contingent contract then it can be implemented by a linear contract. 
\end{theorem}


\section{Conclusion}

Acyclicity conditions are often easily derived as necessary behavioral consequences of various models of utility maximization. It turns out that these simple conditions are remarkably powerful: as shown in this survey, very often they completely characterize the models of interest. This is true, not just of the classical model of consumer utility-maximization, but also of various extensions and variations of that model.


\bibliographystyle{myplainnat}
\bibliography{EconReferences}

\appendix

\end{document}